\documentclass[10pt]{article}
\usepackage{amets,amssymb,amsmath,amstext,lineno,color,xcolor,comment,grffile,array,multirow,makecell,nicefrac}
\usepackage[percent]{overpic}
\usepackage{hyperref,graphicx,multicol,pdfpages}
\hypersetup{colorlinks = true, allcolors = blue, allbordercolors = white}
\usepackage[round,authoryear]{natbib}
\usepackage{float,rotating,epic,tabularx}
\bibpunct[,]{(}{)}{;}{a}{}{,}
\begin{document}
\renewcommand{\baselinestretch}{1.0}\small\normalsize

\small
\noindent
The following is an author-produced version of an article accepted for
publication following peer review (received 3 September 2017; received
in revised form 11 July 2018; accepted 12 July 2018).  The definitive
version from the publisher can be cited as
\href{https://doi.org/10.1016/j.rse.2018.07.016}{Danielson, R. E.,
  J. A. Johannessen, G. D. Quartly, M.-H. Rio, B. Chapron, F. Collard,
  C. Donlon, 2018: Exploitation of error correlation in a large
  analysis validation: GlobCurrent case study.  Remote Sens. Environ.,
  217, 476-490, doi:10.1016/j.rse.2018.07.016.}  This paper
successfully accommodates an early notion of nonlinear signal in a
(noncanonical) measurement model, but does not distinguish between
``agreement'' and ``association''.  Also note that a full model
solution was introduced at OceanPredict 2019 (cf. Fig.~10 poster
update below) and the canonical model (and a distinction between
calibration--agreement versus correlation--association) later appeared
in the appendix and footnotes of
\href{https://doi.org/10.5194/ascmo-6-31-2020}{https://doi.org/10.5194/ascmo-6-31-2020}.

\begin{center}
\vspace{0.2in}
{\Large Exploitation of error correlation in\\
a large analysis validation: GlobCurrent case study\\}
\vspace{0.3in}

\small
\noindent
Richard E. Danielson\textsuperscript{a,*}, Johnny A.
Johannessen\textsuperscript{a}, Graham D. Quartly\textsuperscript{b},\\
Marie-H\'{e}l\`{e}ne Rio\textsuperscript{c}, Bertrand Chapron\textsuperscript{d},
Fabrice Collard\textsuperscript{e}, Craig Donlon\textsuperscript{f}\\
\end{center}
\vspace{0.1in}

\noindent
\textsuperscript{a}Nansen Environmental and Remote Sensing Center, Bergen, Norway\\
\textsuperscript{b}Plymouth Marine Laboratory, Plymouth, United Kingdom\\
\textsuperscript{c}Collecte Localisation Satellites, Ramonville Saint-Agne, France\\
\textsuperscript{d}Ifremer, Plouzan\'{e}, France\\
\textsuperscript{e}OceanDataLab, Locmaria-Plouzan\'{e}, France\\
\textsuperscript{f}European Space Agency, Noordwijk, Netherlands\\
\textsuperscript{*}Danielson Associates Office Inc., Halifax, Canada (rickedanielson@gmail.com)\\

\begin{center}
\vspace{0.3in}

\large
Abstract\\
\end{center}
\normalsize

An assessment of variance in ocean current signal and noise shared by
in situ observations (drifters) and a large gridded analysis
(GlobCurrent) is sought as a function of day of the year for 1993-2015
and across a broad spectrum of current speed.  Regardless of the
division of collocations, it is difficult to claim that any synoptic
assessment can be based on independent observations.  Instead, a
measurement model that departs from ordinary linear regression by
accommodating error correlation is proposed.  The interpretation of
independence is explored by applying Fuller's (1987) concept of
equation and measurement error to a division of error into shared
(correlated) and unshared (uncorrelated) components, respectively.
The resulting division of variance in the new model favours noise.
Ocean current shared (equation) error is of comparable magnitude to
unshared (measurement) error and the latter is, for GlobCurrent and
drifters respectively, comparable to ordinary and reverse linear
regression.  Although signal variance appears to be small, its utility
as a measure of agreement between two variates is highlighted.

Sparse collocations that sample a dense (high resolution) grid permit
a first order autoregressive form of measurement model to be
considered, including parameterizations of analysis-in situ error
cross-correlation and analysis temporal error autocorrelation.  The
former (cross-correlation) is an equation error term that accommodates
error shared by both GlobCurrent and drifters.  The latter
(autocorrelation) facilitates an identification and retrieval of all
model parameters.  Solutions are sought using a prescribed calibration
between GlobCurrent and drifters (by variance matching).  Because the
true current variance of GlobCurrent and drifters is small, signal to
noise ratio is near zero at best.  This is particularly evident for
moderate current speed and for the meridional current component.

\vspace{0.3in}
\noindent\rule{\textwidth}{0.5pt}

\newpage
\begin{multicols}{2}
\normalsize

\section{Introduction}

The idea that errors in two collocated estimates of ocean current
could be independent of each other is, like geostrophy itself, both
practical and instructive.  The difficult implication is that only
signal (or truth) is correlated while noise (or error) is not.
Considering that all measurement models are approximate
\citep{Box_1979}, such a clean separation may be ideal in principle
but is probably quite rare in practice.  The purpose of this study is
to assess the GlobCurrent analysis, but the need to accommodate
cross-correlated errors between GlobCurrent and drifters is not
matched by an existing framework for doing so.  Thus, a new
measurement model is called for.

Although there is no evidence that ocean current signal is dictated by
drifters alone, drifters are employed to refine the mean dynamic
topography (MDT; \citealt{Rio_Hernandez_2004, Rio_etal_2014}).  Thus,
measurement errors may be correlated because the MDT effectively
determines GlobCurrent in a time-mean sense.  Measurement error is not
the only type of error, however.  Perhaps the simplest measurement
models (including all models of this study) assume that truth and
error in a dataset are {\it additive} and the signal in two datasets
can be {\it linearly} related.  There is growing evidence that for
datasets that do not conform exactly to such assumptions, an
associated {\it equation error} term needs to be considered
\citep{Fuller_1987, Carroll_Ruppert_1996, Kipnis_etal_1999}.  It is
precisely because equation error may be strongly correlated that
datasets should not necessarily be considered independent, even if
there is no apparent physical relationship between them.

This study represents an experiment in ocean surface current
validation that draws on advances in measurement modelling, notably in
hydrology and epidemiology, but contemporary surface current
validation also informs this work.  \citet{Johnson_etal_2007}
attribute differences between the OSCAR five-day current analysis and
in situ observations in part to dynamic processes that are difficult
to resolve (e.g., tropical instability waves and high latitude
eddies).  Additionally, although larger signal and noise are resolved
by OSCAR relative to an assimilative model, Johnson et al.~highlight
the existence of intrinsic challenges in capturing the meridional
current near the equator and variability in both components near the
poles.

Surface current validation by \citet{Blockley_etal_2012} and
\citet{Sudre_etal_2013} similarly acknowledge in situ error.  Blockley
et al.~highlight differences in the western equatorial Pacific between
surface currents that they derive from in situ observations and the
FOAM assimilative model.  Global correlation between model and
observations is again much better for the zonal current component
(versus meridional), especially in the tropics and north Pacific
(reduced correlation in the Atlantic is attributed to slightly greater
coverage by eddies).  Although the GECKO satellite-based analysis of
Sudre et al.~finds corresponding systematic variations (by latitude
and current component), their combination of geostrophic and Ekman
estimates is also significantly correlated with in situ estimates.  It
is the agreement between, and independence of, two such estimates that
we wish to reconsider below.

It is convenient to speak of correlation either in terms of signal and
noise, or equivalently, truth and error.  It is also useful to
distinguish between the (spatial or temporal) autocorrelation of a
single variable and the cross-correlation of two variables.
Geophysical modelling approaches (including this study) often assume
that autocorrelation should be easy to find in high resolution
(analysis) data.  For some (in situ) collocation subset, an
affine signal model with additive, orthogonal (or signal-uncorrelated)
noise is often applied.  More formally, if two collocated ocean current
datasets ($I$ and $A$) are divided parsimoniously into shared truth
($t$) and additive error ($\epsilon$) such that
\begin{eqnarray}
  \begin{array}{r} \mathrm{in\ situ}\ \\
                   \mathrm{analysis} \end{array}
  \begin{array}{r} I \\ A \end{array}
  \begin{array}{c} = \\  = \end{array}
  \begin{array}{l} \color{white} \alpha + \beta \color{black} t + \epsilon_I \\
                                 \alpha + \beta               t + \epsilon_A, \end{array}
  \label{eq00}
\end{eqnarray}
then the affine signal model is a linear calibration involving an
unbiased intercept ($\alpha$) and slope ($\beta$) that relates signal
in the two datasets by $A_{signal} = \alpha + \beta I_{signal}$ (where
$I_{signal} = t$).  The measurement model (\ref{eq00}) is known as a
regression model with errors in the variables ($I$ and $A$) but (with
reference to a linear relationship between $I_{signal}$ and
$A_{signal}$) no error in the equation \citep{Fuller_2006}.  Note also
that cross-correlation is only expected from truth, or perhaps error,
that is shared between datasets and that (\ref{eq00}) omits a
partition of error into shared and unshared, or cross-correlated and
uncorrelated, components.

If there is no obvious physical dependence between datasets, then
there is no guarantee that shared error, or shared truth for that
matter, exist.  Because the geophysical interpretation of
cross-correlated error continues to evolve, this concept of sharing is
at least partly unfamiliar, even in the context of two datasets
(\ref{eq00}).  An established explanation in the context of three
datasets \citep{Stoffelen_1998, OCarroll_etal_2008} focuses on the
cross-correlated part of representativeness error: it is natural for
correlation to exist between two higher resolution datasets on scales
that a lower resolution dataset cannot resolve, but if there is a
truth that is shared by all three datasets, then by definition, this
truth is also low resolution and any high resolution correlation must
be considered erroneous, albeit perfectly natural.  Errors of
representation in geophysics (e.g., mismatches that can be written as
a component of $\epsilon_I$ or $\epsilon_A$, as in
\citealt{Gruber_etal_2016a}) refer to information that is beyond some
true, or target, spatiotemporal resolution limit.  However, if shared
truth does exist, it follows that the most generic and inclusive
definition of limitations in this truth is needed to define what
remains in each individual dataset as error.

Stoffelen's introduction of the triple collocation model provides an
important description, and one of the earliest quantifications, of
representativeness error (see also \citealt{Vogelzang_etal_2011}).
Nevertheless, the triple collocation model is just identified, so the
parameters sought (see Appendix) are equal in number to the first and
second moment equations that are available
(cf.~\citealt{Gillard_Iles_2005}).  A familiar characteristic of this
model (like simpler regression models) is its limited flexibility to
identify more parameters.  Hence, correlated representativeness error,
and cross-correlated error in general, must either be known in advance
or perhaps be justifiably small for a retrieval of the triple
collocation parameters.

\citet{Caires_Sterl_2003} discovered a way to explore cross-correlated
error (between altimeters) in comparative applications of the triple
collocation model.  They examined significant wave height and 10-m
wind speed estimates from buoys and two altimeters, which were
carefully averaged to be comparable in space and time with collocated
ERA-40 estimates.  Because representativeness errors were reduced by
this averaging, it was postulated that any remaining ERA-40
cross-correlated errors could be neglected if ERA-40 did not
assimilate an observational dataset.  A bound on cross-correlated
error was then estimated for the altimeters, whose uncorrelated error
was found to be relatively low when retrieved together with ERA-40
rather than separately with ERA-40 and buoys.  Consideration of this
bound yielded an increase in altimeter error variance by a factor of
two or more, but Caires and Sterl suggested that cross-correlated
error may have been smaller.

\citet{Janssen_etal_2007} examined wave height data from two
altimeters, buoys, and an ECMWF wave hindcast, and employed an
iterative form of orthogonal regression \citep{Gillard_Iles_2005} with
estimates of uncorrelated error from the triple collocation model.  An
important acknowledgement was given of the linear calibration in
(\ref{eq00}) being a potential source of cross-correlated error (i.e.,
where a nonlinear signal model might be appropriate instead).  As in
\citet{Caires_Sterl_2003}, it was postulated that cross-correlated
errors could be neglected if data (or systematic errors) were not
assimilated, but uncorrelated altimetric error was again found to be
relatively low when the triple collocation model was applied to both
altimeters at once.  Janssen et al.\ proposed additional model
equations (using ECMWF first guess and analysis wave products) to
quantify rather than just bound most errors, but found that altimetric
error, including its cross-correlated component, was small.

Methods of collocating buoy, radiometer, and microwave SST estimates
(e.g., \citealt{OCarroll_etal_2008}) also point to cross-correlated
error being small, but only insofar as representativeness error is
tested, as above, by parameter comparisons.  A novel assessment of
cross-correlated error has also been given using a high resolution,
rescaled in situ dataset as a proxy for truth.
\citet{Yilmaz_Crow_2014} use this proxy to directly characterize terms
of the triple collocation model based on soil moisture from an
assimilative model and soil moisture retrievals from passive (AMSR-E)
and active (ASCAT) satellites.  The dependence of satellite retrievals
is notable because significant cross-correlated errors are found.
This study concludes that zero error cross-correlation is a tenuous
assumption of the triple collocation model as its corresponding bias
in parameter retrievals is systematic.

Contemporary calibration and validation studies have introduced a
growing list of geophysical dataset differences, which taken together,
define corresponding limitations on shared truth.  However, perhaps
the most generic characterization of these limitations is found in the
measurement modelling literature: \citet{Fuller_1987} defines
measurement error in the familiar sense of random data departures from
a linear regression solution and distinguishes {\it equation error} as
random departures from the linear signal model of (\ref{eq00}), owing
to nonlinearity in the signal model of interest.
\citet{Carroll_Ruppert_1996} expose the importance of this refinement
in a geophysical application and, as noted above,
\citet{Janssen_etal_2007} highlight that such nonlinearity is a
potential source of cross-correlated error.

The combination of measurement error and equation error is useful to
better accommodate limitations in the scope of a shared truth.  With
reference to person-specific bias in epidemiology,
\citet{Kipnis_etal_1999, Kipnis_etal_2002} introduce equation error as
two additional terms ($\epsilon_{QI}$ and $\epsilon_{QA}$) in
(\ref{eq00}) that lead to
\begin{eqnarray}
  \begin{array}{r} \mathrm{in\ situ}\ \\
                   \mathrm{analysis} \end{array}
  \begin{array}{r} I \\ A \end{array}
  \begin{array}{c} = \\  = \end{array}
  \begin{array}{l} \color{white} \alpha + \beta \color{black} t + \epsilon_{QI} + \epsilon_I \\
                                 \alpha + \beta               t + \epsilon_{QA} + \epsilon_A, \end{array}
  \label{eqerr}
\end{eqnarray}
where $\epsilon_I$ and $\epsilon_A$ are now random departures from a
possibly nonlinear signal model.
\citet{Carroll_Ruppert_1996} note that applications of (\ref{eqerr})
have been limited, possibly because if $\epsilon_{QI}$ and
$\epsilon_{QA}$ are considered to be independent of other errors, they
can be recombined with $\epsilon_I$ and $\epsilon_A$ to yield the
simpler equation (\ref{eq00}) with its original properties intact
\citep{Moberg_Brattstrom_2011}.  Below, the same linear signal model
as in (\ref{eq00}) will be considered, with shared equation error
defined by $\epsilon_{QI} = \epsilon_{QA}$ and total error involving
both shared and unshared components.  In other words, equation error
is not independent so it is important to quantify this as a separate
term in our application of (\ref{eqerr}).



In addition to the interpretation of cross-correlated errors, there
remains the issue of identifying solutions to increasingly
sophisticated statistical models.  Increasing the number of collocated
datasets (e.g., \citealt{Janssen_etal_2007, Zwieback_etal_2012,
  Gruber_etal_2016b}) is one approach.  However, an important
development in the geophysical literature is the recognition by
\citet{Su_etal_2014} that three or more datasets may be unnecessary,
as collocation models appear to belong to a broader family of
instrumental variable regression models, and within this family, a
precedent exists for using lagged variables as instruments.  Following
Su et al., this implies that by embracing autocorrelation, strategies
should continue to emerge that depend on fewer datasets to identify a
larger number of collocations and statistical model parameters.  By
comparison with the error-in-variables model (\ref{eq00}), the novelty
of the strategy proposed below is that it also permits the retrieval
of variance in shared error and, in one ocean current experiment, also
equation error.

The present study seeks to advance measurement modelling and parameter
identification with the benefit of error correlation.  The focus is on
ocean surface current validation, but general supporting concepts and
terms (such as {\it measurement model}) are provided in the Appendix.
The next section describes the collocation of GlobCurrent and drifters
and proposes a commonly prescribed linear relationship between them
that addresses the difference in variance between these two datasets.
Formulation of a measurement model that permits error correlation to
be exploited is given in Section~3.  We then describe the strong and
weak constraints that allow a retrieval of all model parameters and
assess the performance of GlobCurrent and drifter data in Section~4.
Throughout this paper, equal emphasis is placed on true variance and
on the contributions to total error.  Discussion of inferences based
on the division of variance into shared truth and error are
highlighted in Section~5 and Section~6 contains the conclusions.
\end{multicols}

\begin{figure}[htb]
  \centering
  \includegraphics[width=0.75\textwidth]{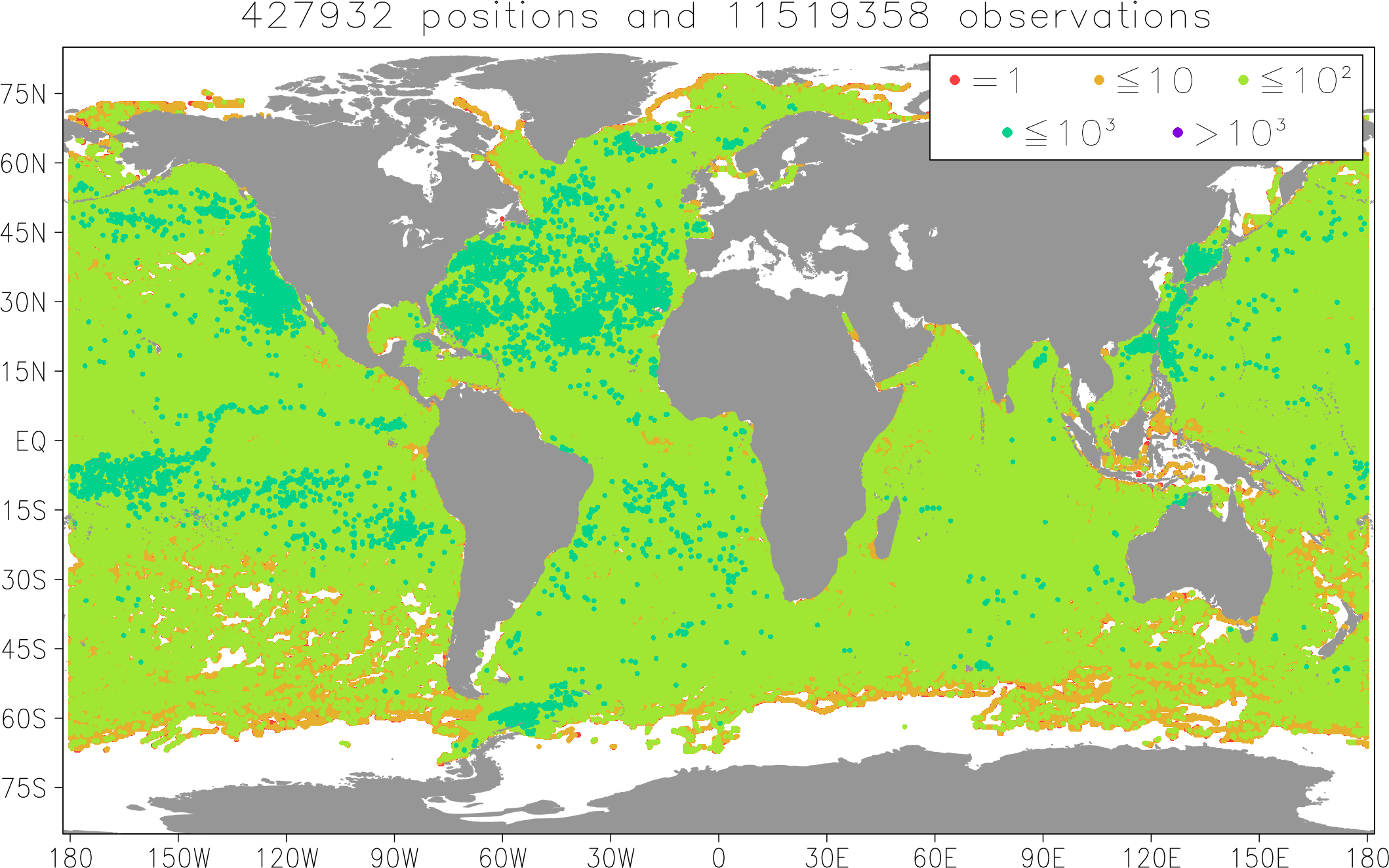} 
  \caption{Number of surface drifter velocity observations between
    January 1993 and December 2015 (order of magnitude in colour) with
    drogues attached.  Shown are values at the $\nicefrac{1}{4}^\circ$ resolution
    of the GlobCurrent grid (i.e., collocations are nearest
    neighbours).}
  \label{fig01}
\end{figure}

\begin{multicols}{2}

\section{Selection of a calibration}

We begin with the idea that GlobCurrent and drifters provide estimates
of fundamentally different ocean currents, but they also provide
overlapping views of a true (or target) ocean current that can be
represented at 15~m below the surface on a 6-h,
$\nicefrac{1}{4}^\circ$ grid.  By any definition of shared truth, both
GlobCurrent and drifters have errors.  GlobCurrent is an analysis that
linearly combines the geostrophic and Ekman components.  Drifters
respond locally to a combination of geostrophic, Ekman, tidal,
inertial, Stokes, and wind drift processes, including (erroneous)
processes on scales smaller and faster than the GlobCurrent grid can
resolve.  In general, such differences can be considered a mismatch in
their supports (see Appendix).  Nearest-neighbour collocations of
drifters (whose drogues move roughly with the 15-m current) and
GlobCurrent (also at 15~m, with additional samples at daily intervals)
are considered below.

Six-hourly drifter velocity has been estimated following
\citet{Hansen_Poulain_1996}.  We restrict attention to drifters whose
continuous drogue presence was confirmed by objective or subjective
means \citep{Rio_2012, Lumpkin_etal_2013}.  The resulting geographic
distribution for 1993-2015 (Fig.~\ref{fig01}) yields more than eleven
million drifter and GlobCurrent zonal and meridional velocity
estimates (\citealt{Danielson_2017}; a comparable number of drifters
lost their drogues and, being more responsive to surface wind forcing,
are ignored).  It is convenient to divide collocations by even and odd
year, with the latter subset permitting an independent check on
calculations.  Below, only the even-year subset is discussed but the
same conclusions can be obtained from the results (available as
supplementary material) of the odd-year subset.
\end{multicols}

\begin{figure}[htb]
  \centering
  \includegraphics[width=0.85\textwidth]{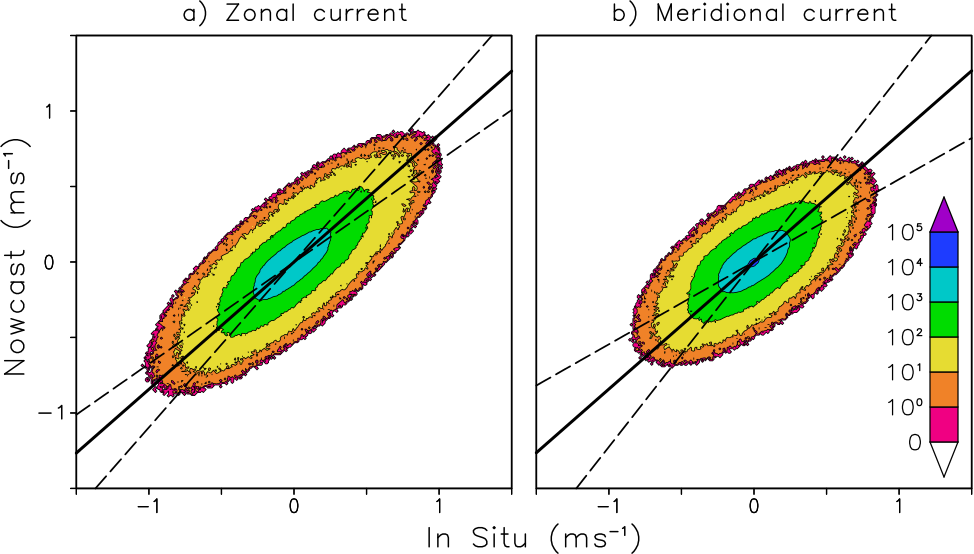}
  \caption{Two-dimensional histograms of a) zonal and b) meridional
    15-m current component for 5310226 non-outlier collocations from the even
    years between 1993 and 2015 (approximately half the collocations
    of Fig.~\ref{fig01}, after removing about 10\% of these data as
    outliers following \citealt{Hubert_etal_2012}).  The dashed lines
    are the ordinary (shallow slope) and reverse (steep slope) linear
    regression references for each current component.  The slope of
    the solid line is defined by the GlobCurrent--drifter
    variance ratio (the same ratio for both current components; see
    next section).  The logarithmic colourbar is number of values in
    0.01-ms$^{-1}$ bins.}
  \label{fig02a}
\end{figure}

\begin{multicols}{2}
Joint frequency of occurrence by current component, including the full
range of possible linear calibrations of GlobCurrent relative to
drifters, is shown in Fig.~\ref{fig02a}.  These two-dimensional
histograms are rather well behaved following removal of about 10\% of
the most extreme current values \citep{Hubert_etal_2012}.  Similar
regression slopes are revealed in both the zonal and meridional
distributions.  Between the bounding ordinary and reverse linear
regression reference slopes (dashed lines) is a slope defined by the
ratio of total variance between GlobCurrent and drifters (solid line;
defined in the next section).  Unfortunately, scatter away from these
regression lines is a poor indication that there might be a component
of error variance that is shared between GlobCurrent and drifters, or
that total error variance might be greater than the variance in shared
truth.

\begin{figure}[htb]
  \centering
  \hbox{
    \begin{overpic}[height=0.4\textwidth]{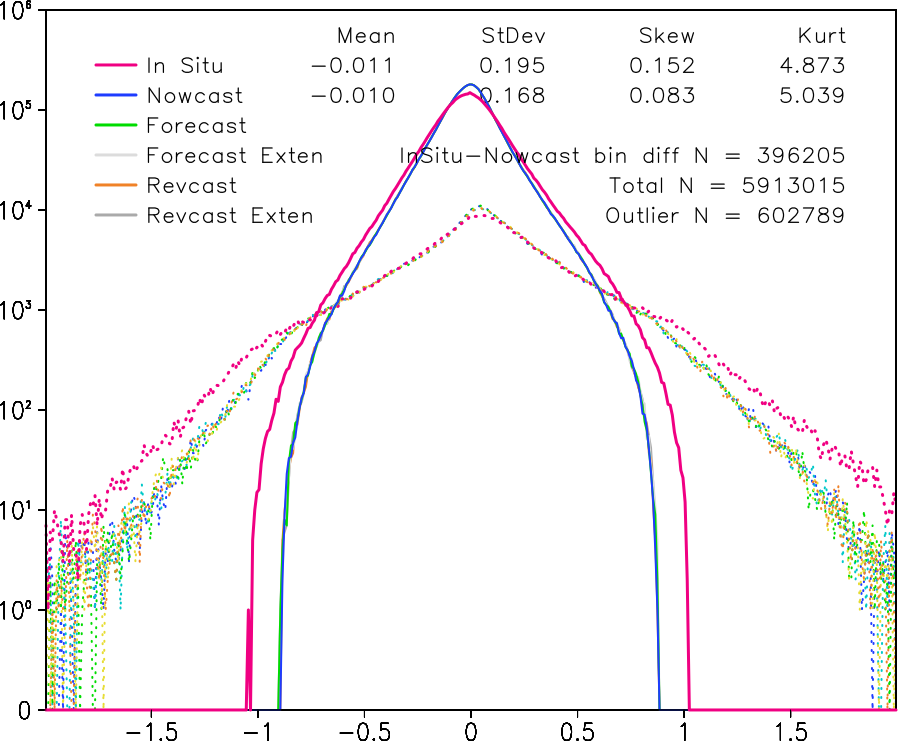} 
      \put (33,85){\footnotesize a) Zonal current}
    \end{overpic}
    \begin{overpic}[height=0.4\textwidth]{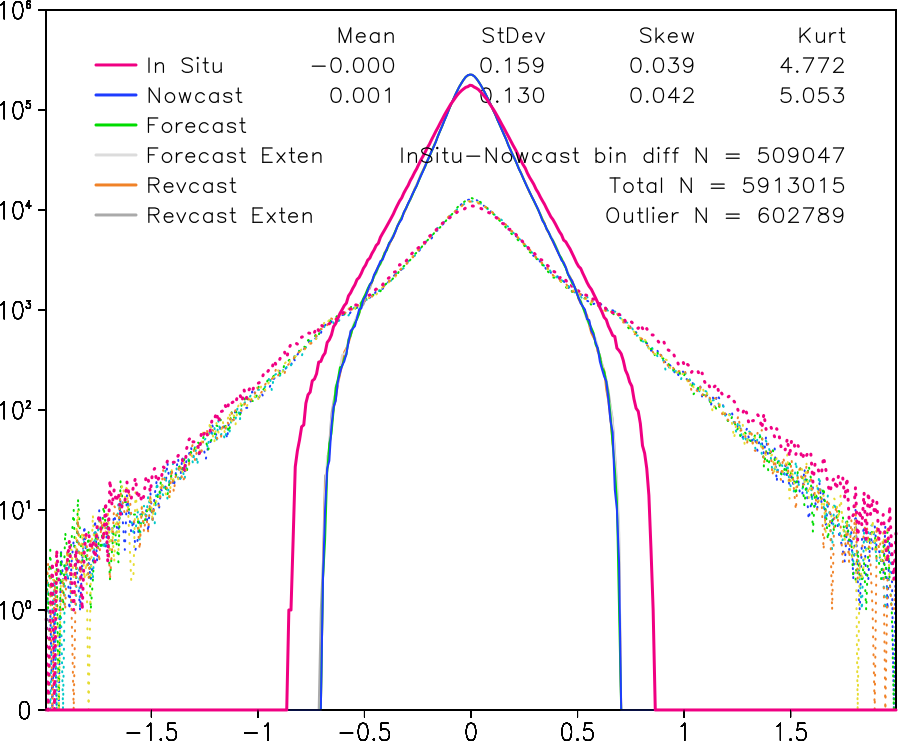} 
      \put (25,85){\footnotesize b) Meridional current}
      \put (-45,-7){\footnotesize Ocean current magnitude at 15-m depth (ms$^{-1}$)}
    \end{overpic}
  }
  \vspace{0.1in}
  \caption{One-dimensional histograms of a) zonal and b) meridional
    15-m current component, as in Fig.~\ref{fig02a}, but including
    outliers separately (dotted lines).  Also shown are drifter (red)
    and GlobCurrent nowcast (blue), forecast (green and light grey),
    and revcast (orange and dark grey) histograms.  Forecast and
    revcast data are taken one day (with extended data from two days)
    before and after each collocation, respectively.  Statistical
    moments of the non-outlier in situ and nowcast distributions are
    given with a measure of difference between the two (i.e., one half
    of the in situ minus nowcast bin count difference).  The
    logarithmic ordinate is number of values in 0.01-ms$^{-1}$ bins.}
  \label{fig02}
\end{figure}

The corresponding one-dimensional (marginal) distributions
(Fig.~\ref{fig02}) highlight an unsurprising difference between
current estimates: because drifters capture a greater range of
physical processes at higher resolution, we find fewer values near zero and
more values of large magnitude than GlobCurrent (with an equal number at about
$\pm$0.15 ms$^{-1}$).  Also as expected, GlobCurrent samples at two
days (extended forecast) and one day (forecast) before each drifter
(in situ) observation, as well as one day (revcast) and two days
(extended revcast) after, have the same distribution as the
GlobCurrent collocations (nowcast).  Outliers are shown separately by
dotted lines in Fig.~\ref{fig02} and are identified by minimizing the
covariance matrix determinant for the six estimates of zonal and
meridional current \citep{Hubert_etal_2012}.  Because covariance (and
skewness and kurtosis) are sensitive to outliers
\citep{McColl_etal_2014, Su_etal_2014}, collocation groups are trimmed
by about 10\% before other calculations are performed.  Often this
excludes extreme values in the zonal or meridional component and
values close to zero in the opposite component.
\end{multicols}

\begin{multicols}{2}
The distinction between cross-correlated and uncorrelated error is
sufficiently novel that initial solutions of (\ref{eqerr}) benefit
from the assumption of a fixed calibration that can be applied
uniformly.  (Subsequent work will seek a general, varying solution,
but this simplification applies to all experiments below.)  An
assumption that would be consistent with the mismatch in GlobCurrent
and drifter {\it support} (rather than a bias between them) is that
both are already unbiased.  However, we note in Section~4 that if
calibration is bounded by ordinary and reverse linear regression
(dashed lines in Fig.~\ref{fig02a}), then this assumption would not
apply to all collocation subsets.  An alternate assumption that can be
applied uniformly, and whose bias is familiar in the context of
(\ref{eq00}), is known as variance matching \citep{Fuller_2006,
  Yilmaz_Crow_2013, Su_etal_2014}.  This calibration is marked by a
lack of assumptions about relative error in GlobCurrent and drifters.
It fixes regression slope midway between the bounding ordinary and
reverse linear regression solutions (solid line in Fig.~\ref{fig02a})
and fixes GlobCurrent and drifter signal-to-noise ratio (SNR) to be
equal.  A definition and further implications are given in Section~3.

\begin{figure}[htb]
  \centering
  \includegraphics[width=0.85\textwidth]{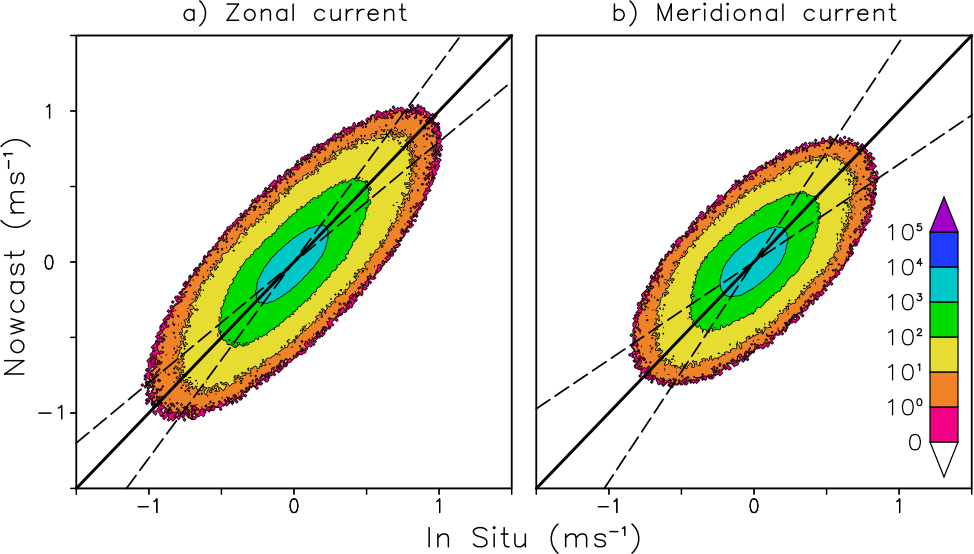}
  \caption{As in Fig.~\ref{fig02a}, but after dividing all GlobCurrent
    data by 0.84 (i.e., the ratio of nowcast to drifter standard
    deviation), where zonal and meridional components are expressed as
    complex numbers and the same variance match is applied to both
    components.}
  \label{fig03a}
\end{figure}

\begin{figure}[htb]
  \vspace{0.1cm}
  \centering
  \hbox{
    \begin{overpic}[height=0.4\textwidth]{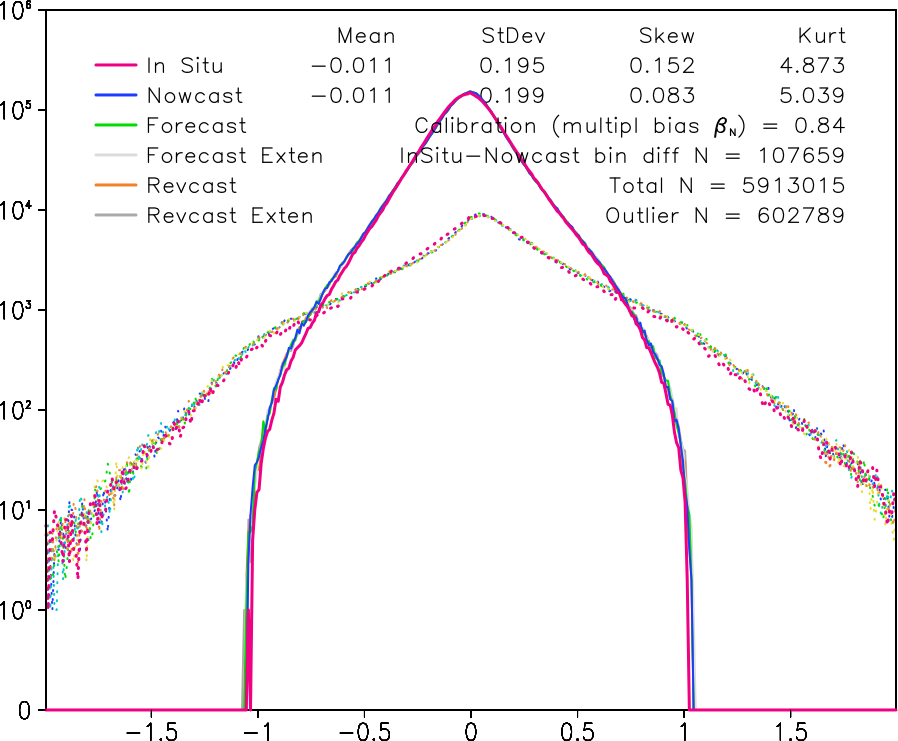} 
      \put (33,85){\footnotesize a) Zonal current}
    \end{overpic}
    \begin{overpic}[height=0.4\textwidth]{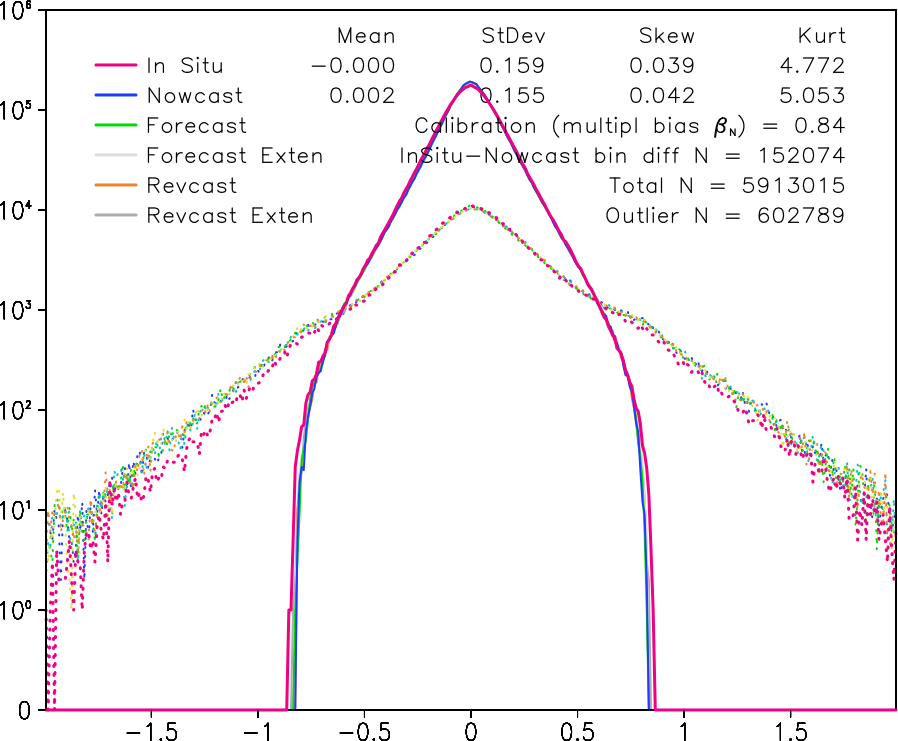} 
      \put (25,85){\footnotesize b) Meridional current}
      \put (-45,-7){\footnotesize Ocean current speed at 15-m depth (ms$^{-1}$)}
    \end{overpic}
  }
  \vspace{0.1in}
  \caption{As in Fig.~\ref{fig02}, but after dividing all GlobCurrent
    data by 0.84.}
  \label{fig03}
\end{figure}

Figures~\ref{fig03a} and \ref{fig03} are the result of matching the
variance of GlobCurrent to that of drifters.  (Simultaneous matching
of the zonal and meridional components is accomplished by expressing
these two components as a complex number.)  Dividing the GlobCurrent
data by a standard deviation ratio of 0.84 reduces the number of values
near zero and increases the number of large magnitude values, as expected.
This calibration removes much of the cumulative difference in bin counts:
from 7-8\% in Fig.~\ref{fig02} to about 2\% in Fig.~\ref{fig03}.
However, the distinction between calibrated GlobCurrent and drifters
remains, as histogram shape is otherwise preserved (note that skewness
and kurtosis are variance-normalized moments) and current direction is
unchanged.  Moreover, and notwithstanding important applications to
assimilation and model validation (e.g., \citealt{Stoffelen_1998,
  Tolman_1998}), this distinction would remain at least under any
affine calibration.
\end{multicols}

\newpage
\begin{multicols}{2}

\section{Measurement model development}

A series of experimental models, based initially on the triple
collocation approach \citep{Stoffelen_1998, McColl_etal_2014} with
solutions sought by the method of moments \citep{Gillard_Iles_2005},
have informed the measurement model that we will focus on.  The first
experimental model in this series (\ref{eq01}) can be criticised for
using extrapolated (forecast and revcast) GlobCurrent estimates
assuming that extrapolated errors are independent.  Gridded altimetric
data are often based on a centered span of up to 12~days of
Topex/Jason passes and a longer period for Envisat.  Similarly for the
Ekman (or Stokes) current estimates from a model-based analysis, if a
model has the wind front in the wrong location or an incorrect initial
storm intensity, it may retain a consistent bias for days.  Thus, the
assumption of independent errors $\epsilon$ in a slightly modified
triple collocation model,
\begin{eqnarray}
  \begin{array}{r} \mathrm{in\ situ}\ \\
                   \mathrm{forecast} \\
                   \mathrm{revcast} \end{array}
  \begin{array}{r} I \\ F \\ R \end{array}
  \begin{array}{c} = \\  =  \\ = \end{array}
  \begin{array}{l} \color{white} \alpha_F + \beta_F \color{black} t + \epsilon_I \\
                   \alpha_F + \beta_F t + \epsilon_F \\
                   \alpha_R + \beta_R t + \epsilon_R, \end{array}
  \label{eq01}
\end{eqnarray}
can be considered experimental at best.  Note that $\alpha$, $\beta$,
$t$, and $\epsilon$ are additive calibration, multiplicative
calibration (or regression slope), truth, and error, respectively, and
our use of drifters as a calibration reference implies that $\alpha_I
= 0$ and $\beta_I = 1$.  Here, $F$ and $R$ are obtained by extrapolation
of GlobCurrent from outside a centered window of only a few days.

The form of (\ref{eq01}) is recognizable in an intermediate (but still
unsatisfactory) model (\ref{eq02}) that includes both GlobCurrent and
drifter collocations ($I$ and $N$) and retains additive and
multiplicative calibration parameters ($\alpha$ and $\beta$) for each
GlobCurrent estimate.  A notable simplification of (\ref{eq02}) is
that extrapolation is replaced by a persistence forecast/revcast, so
$F$ and $R$ are just GlobCurrent samples taken one day before and
after each collocation, respectively.
\begin{eqnarray}
  \begin{array}{r} \mathrm{in\ situ}\ \\
                   \mathrm{nowcast} \\
                   \mathrm{forecast} \\
                   \mathrm{revcast} \end{array}
  \begin{array}{r} I \\ N \\ F \\ R \end{array}
  \begin{array}{c} = \\  =  \\ = \\ = \end{array}
  \begin{array}{l} \color{white}{\alpha_N + \beta_N} \color{black} t + \epsilon_I \\
                                 \alpha_N + \beta_N                t + \epsilon_N \\
                                 \alpha_F + \beta_F                t + \epsilon_N + \epsilon_F \\
                                 \alpha_R + \beta_R                t + \epsilon_N + \epsilon_R. \end{array}
  \label{eq02}
\end{eqnarray}
The model (\ref{eq02}) is overly constrained in its treatment of
correlated error, however.  There is no shared (equation) error
between GlobCurrent and drifters and a complete sharing of $N$ errors
in $F$ and $R$.  In turn, it is perhaps unsurprising that there may be
effectively no difference (in terms of physical insight) between
parameter retrievals based on (\ref{eq02}) and ordinary and reverse
linear regression references based on $I$ and $N$ alone
\citep{Danielson_etal_2017}.

Two further innovations are required to arrive at the measurement
model of interest.  One is that a first-order autoregressive (AR-1)
parameterization is probably the simplest way to accommodate both
GlobCurrent-drifter error cross-correlation as well as GlobCurrent
error autocorrelation.  Error propagation is parameterized in the same
sense as it might occur in an ocean current analysis, with
observational error having its biggest impact on an analysis at the
time of observation, with a decreasing, but symmetric impact at times
before and after.  The AR-1 form accommodates autocorrelated errors
(e.g., from altimetry) that also have a symmetric upstream and
downstream impact (note that asymmetric error propagation may be
appropriate in some applications).

The second innovation, following \citet{Su_etal_2014}, is that
additional, or extended, samples of GlobCurrent are beneficial,
assuming these remain inside the autocorrelation envelope.  The
resulting model becomes
\end{multicols}
\vspace{-0.3in}
\begin{eqnarray}
  \begin{array}{r} \mathrm{in\ situ}\ \\
                   \mathrm{nowcast} \\
                   \mathrm{forecast} \\
                   \mathrm{extended\ forecast} \\
                   \mathrm{revcast} \\
                   \mathrm{extended\ revcast} \end{array}
  \begin{array}{r} I \\ N \\ F \\ E \\ R \\ S \end{array}
  \begin{array}{c} = \\  =  \\ = \\ = \\ = \\ = \end{array}
  \begin{array}{l} \color{white}{\alpha_N + \beta_N} \color{black} t + \color{white}{\lambda_E (  \lambda_F ( \lambda_N} \color{black} \epsilon_I \\
                                 \alpha_N + \beta_N  t + \color{white}{\lambda_E (                \lambda_F (} \color{black} \lambda_N \epsilon_I + \epsilon_N \\
                                 \alpha_F + \beta_F  t + \color{white}{\lambda_E (} \color{black} \lambda_F ( \lambda_N \epsilon_I + \epsilon_N ) + \epsilon_F \\
                                 \alpha_E + \beta_E  t +               \lambda_E (                \lambda_F ( \lambda_N \epsilon_I + \epsilon_N ) + \epsilon_F ) + \epsilon_E \\
                                 \alpha_R + \beta_R  t + \color{white}{\lambda_E (} \color{black} \lambda_R ( \lambda_N \epsilon_I + \epsilon_N ) + \epsilon_R \\
                                 \alpha_S + \beta_S  t +               \lambda_S (                \lambda_R ( \lambda_N \epsilon_I + \epsilon_N ) + \epsilon_R ) + \epsilon_S, \end{array}
  \label{infers}
\end{eqnarray}
\vspace{-0.2in}

\begin{multicols}{2}
\noindent
where Fuller's (1987) equation error, corresponding in (\ref{eqerr})
to $\epsilon_{QI} = \epsilon_{QA}$ \citep{Kipnis_etal_1999}, is the
shared (cross-correlated) error parameterization $\lambda_N
\epsilon_I$.  We return to the interpretation of shared and unshared
error in $\epsilon_I$ below.  The remaining errors are uncorrelated
measurement errors, also denoted individual errors: $\epsilon_N$,
$\epsilon_F$, $\epsilon_E$, $\epsilon_R$, and $\epsilon_S$.

A so-called INFR model, whose name is taken from the data samples on
the LHS of (\ref{eq02}) but whose RHS is taken from (\ref{infers}),
has parameters that are almost identifiable (in a statistical sense).
That is, one can derive 10 covariance equations (given below) but
there are 11 unknown parameters.  The INFERS model (\ref{infers})
includes an extended forecast and revcast, which are GlobCurrent
samples two days before and after each collocation.  Under the
assumption that GlobCurrent errors remain correlated at least over
five days (e.g., as gauged by the product $\lambda_F \lambda_E
\lambda_R \lambda_S$), INFERS is more attractive because there are
more covariance equations (21) than unknown parameters (17).  (Of
course, with more samples further improvement in the ratio of these
numbers is possible.)  Standard assumptions of no correlation between
truth and error (orthogonality) and among individual errors then allow
all elements of the covariance matrix to be defined by
\end{multicols}
\vspace{-0.3in}
\begin{eqnarray}
  \begin{array}{r} Var(I) \\ Var(N) \\ Var(F) \\ Var(E) \\ Var(R) \\ Var(S) \\ Cov(I,N) \\ Cov(I,F) \\ Cov(I,E) \\ Cov(I,R) \\ Cov(I,S) \\ Cov(N,F) \\ Cov(N,E) \\ Cov(N,R) \\ Cov(N,S) \end{array}
  \begin{array}{c}   =    \\   =    \\   =    \\   =    \\   =    \\   =    \\    =     \\    =     \\    =     \\    =     \\    =     \\    =     \\    =     \\    =     \\    =     \end{array}
  \begin{array}{l} \sigma_t^2 + \sigma_I^2 \\ \beta_N^2 \sigma_t^2 + \lambda_N^2 \sigma_I^2 + \sigma_N^2 \\ \beta_F^2 \sigma_t^2 + \lambda_F^2 \lambda_N^2 \sigma_I^2 + \lambda_F^2 \sigma_N^2 + \sigma_F^2 \\ \beta_E^2 \sigma_t^2 + \lambda_E^2 \lambda_F^2 \lambda_N^2 \sigma_I^2 + \lambda_E^2 \lambda_F^2 \sigma_N^2 + \lambda_E^2 \sigma_F^2 + \sigma_E^2 \\ \beta_R^2 \sigma_t^2 + \lambda_R^2 \lambda_N^2 \sigma_I^2 + \lambda_R^2 \sigma_N^2 + \sigma_R^2 \\ \beta_S^2 \sigma_t^2 + \lambda_S^2 \lambda_R^2 \lambda_N^2 \sigma_I^2 + \lambda_S^2 \lambda_R^2 \sigma_N^2 + \lambda_S^2 \sigma_R^2 + \sigma_S^2 \\ \beta_N \sigma_t^2 + \lambda_N \sigma_I^2 \\ \beta_F \sigma_t^2 + \lambda_F \lambda_N \sigma_I^2 \\ \beta_E \sigma_t^2 + \lambda_E \lambda_F \lambda_N \sigma_I^2 \\ \beta_R \sigma_t^2 + \lambda_R \lambda_N \sigma_I^2 \\ \beta_S \sigma_t^2 + \lambda_S \lambda_R \lambda_N \sigma_I^2 \\ \beta_N \beta_F \sigma_t^2 + \lambda_F \lambda_N^2 \sigma_I^2 + \lambda_F \sigma_N^2 \\ \beta_N \beta_E \sigma_t^2 + \lambda_E \lambda_F \lambda_N^2 \sigma_I^2 + \lambda_E \lambda_F \sigma_N^2 \\ \beta_N \beta_R \sigma_t^2 + \lambda_R \lambda_N^2 \sigma_I^2 + \lambda_R \sigma_N^2 \\ \beta_N \beta_S \sigma_t^2 + \lambda_S \lambda_R \lambda_N^2 \sigma_I^2 + \lambda_S \lambda_R \sigma_N^2, \end{array}
  \label{infers_strong}
\end{eqnarray}
\vspace{-0.1in}
and
\vspace{-0.2in}
\begin{eqnarray}
  \begin{array}{r} Cov(F,E) \\ Cov(F,R) \\ Cov(F,S) \\ Cov(E,R) \\ Cov(E,S) \\ Cov(R,S) \end{array}
  \begin{array}{c}    =     \\    =     \\    =     \\    =     \\    =     \\    =     \end{array}
  \begin{array}{l} \beta_F \beta_E \sigma_t^2 + \lambda_E \lambda_F^2 \lambda_N^2 \sigma_I^2 + \lambda_E \lambda_F^2 \sigma_N^2 + \lambda_E \sigma_F^2 \\ \beta_F \beta_R \sigma_t^2 + \lambda_F \lambda_R \lambda_N^2 \sigma_I^2 + \lambda_F \lambda_R \sigma_N^2 \\ \beta_F \beta_S \sigma_t^2 + \lambda_F \lambda_S \lambda_R \lambda_N^2 \sigma_I^2 + \lambda_F \lambda_S \lambda_R \sigma_N^2 \\ \beta_E \beta_R \sigma_t^2 + \lambda_E \lambda_F \lambda_R \lambda_N^2 \sigma_I^2 + \lambda_E \lambda_F \lambda_R \sigma_N^2 \\ \beta_E \beta_S \sigma_t^2 + \lambda_E \lambda_F \lambda_S \lambda_R \lambda_N^2 \sigma_I^2 + \lambda_E \lambda_F \lambda_S \lambda_R \sigma_N^2 \\ \beta_R \beta_S \sigma_t^2 + \lambda_S \lambda_R^2 \lambda_N^2 \sigma_I^2 + \lambda_S \lambda_R^2 \sigma_N^2 + \lambda_S \sigma_R^2. \end{array}
  \label{infers_autocov}
\end{eqnarray}
\vspace{-0.2in}

\begin{multicols}{2}
The corresponding 17 unknowns are true variance ($\sigma_t^2$),
multiplicative calibration for five datasets ($\beta_N, \beta_F,
\beta_E, \beta_R, \beta_S$), and error variance for all six
($\sigma_I^2, \sigma_N^2, \sigma_F^2, \sigma_E^2, \sigma_R^2,
\sigma_S^2$).  There are also five parameters that gauge
GlobCurrent-drifter error cross-correlation ($\lambda_N$ is denoted
shared error fraction below) and GlobCurrent error autocorrelation
($\lambda_F, \lambda_E, \lambda_R, \lambda_S$).  An analytic solution
of all parameters except $\sigma_t^2$ and $\beta_N$ is possible using
(\ref{infers_strong}) as a strong constraint (i.e., using all
variances and the covariances involving the GlobCurrent and drifter
collocations $I$ and $N$).  The remaining equations
(\ref{infers_autocov}) are denoted the autocovariance equations (i.e.,
covariances involving only GlobCurrent forecast and revcast samples
$FERS$).

True variance ($\sigma_t^2$) and multiplicative calibration or
regression slope ($\beta_N$) between GlobCurrent and drifters are key
measurement model parameters.  In the context of INFERS, these are
both free parameters that can be sought numerically using the
autocovariance equations as a weak constraint, that is, by approaching
minima in the difference between the LHS and RHS of
(\ref{infers_autocov}).  Matching GlobCurrent variance to that of
drifters (as in Section~2) provides all experiments with a fixed, but
approximate, slope parameter $\beta_N$.  In other words, our focus on
a search for true variance is also limited by this assumption.  It is
important to note, moreover, that variance matching provides more
freedom to retrieve large cross-correlated error because it is
between the bounding ordinary and reverse linear regression solutions
(i.e., where all variance in either GlobCurrent or drifters is assigned
to truth and the possibility of cross-correlated error is excluded).
It follows from this assumption that
\end{multicols}
\vspace{-0.2in}
\begin{equation}
  \beta_N^2 = Var(N) / Var(I) \hspace{0.5cm} \Rightarrow \hspace{0.5cm} \sigma_N^2 = \sigma_I^2 (\beta_N^2 - \lambda_N^2).
  \label{match}
\end{equation}
\vspace{-0.3in}

\begin{multicols}{2}
The remaining INFERS model parameters are retrieved once a solution
for $\sigma_t^2$ is obtained.  The weakly constrained minimization of
(\ref{infers_autocov}) is sought between bounds for $\sigma_t^2$ that
are given by $Var(I) = \sigma_t^2 + \sigma_I^2$ (i.e., between
$\sigma_t^2 = 0$ and the ordinary linear regression solution of
$\sigma_I^2 = 0$), with the additional strong constraint that all
other variances ($\sigma_N^2, \sigma_F^2, \sigma_E^2, \sigma_R^2,
\sigma_S^2$) also remain non-negative.  Just like the variance matched
solution for $\beta_N$, each zonal and meridional current component is
first expressed as a complex number so that 17 parameters are
identified for both components at the same time (i.e., the covariances
in (\ref{infers_strong}) and (\ref{infers_autocov}) are real parts).

The remainder of this study is a diagnostic exploration of the
parameters obtained from (\ref{infers})-(\ref{match}) given surface
current variations that are jointly sampled by GlobCurrent and
drifters.  As required by INFERS, we also perform a simple check that
GlobCurrent samples of truth and error (combined) remain inside their
autocorrelation envelope: for any group of collocations, the minimum
correlation between an NFERS pair (i.e., between E and S) is expected
to be larger than about 0.7.  All correlation estimates are obtained
from the LHS of (\ref{infers_strong}) and (\ref{infers_autocov}).

\section{Performance assessment}

We introduce a retrieval of measurement model parameters for all
5310226 non-outlier collocations from the even years between 1993 and
2015.  This is followed by retrievals for subsets of this group as a
function of day of the year and current speed.  GlobCurrent and
drifters appear to provide complementary information about ocean
surface current.  The SNR is near zero at best as variance in a shared
true current tends to be smaller than the variance in total (shared
and unshared) error.  We also show that shared error fraction
($\lambda_N$) is quite high.
A posteriori, this motivates our accommodation of cross-correlated
error in (\ref{infers}).  To the extent that cross-correlated error
and equation error are the same (see Section~5), an important question
is raised of whether a linear signal model and additive errors for
GlobCurrent and drifters can be considered robust (and by what
metric).  Large individual (measurement) error is consistent with
GlobCurrent and drifters as offering quite noisy estimates of shared
true current variability (again subject to a linear calibration).  In
Section~5, we find that individual error is similar to the ordinary
and reverse linear regression reference solutions.  In other words, it
is mainly a quantification of shared/correlated truth and error that
require our attention.
\end{multicols}

\begin{figure}[htbp]
  \centering
  \includegraphics[width=0.7\textwidth]{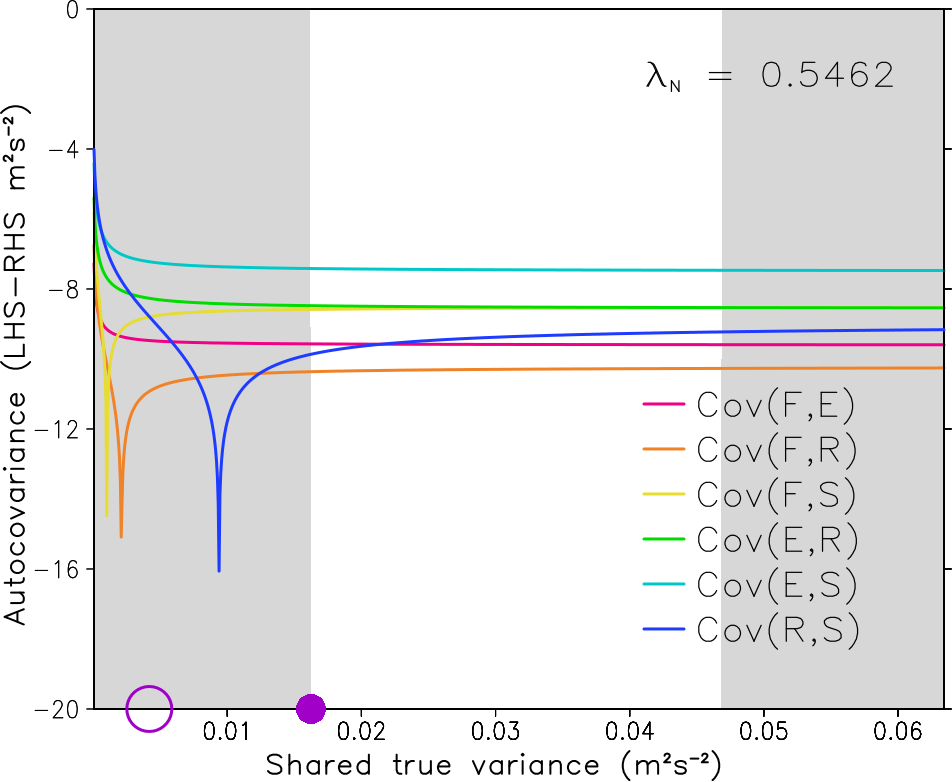} 
  \caption{First demonstration of an INFERS parameter solution by
    weakly constrained minimization of the magnitude of differences
    between the LHS and RHS of the autocovariance equations
    (\ref{infers_autocov}) for the 5310226 non-outlier collocations
    from even years between 1993 and 2015 (roughly half of
    Fig.~\ref{fig02}).  The abscissa is true variance ($\sigma_t^2$)
    in m$^2$s$^{-2}$ between zero and $Var(I)$.  The ordinate is log
    of absolute difference (LHS minus RHS).  Grey shading denotes
    regions of negative error variance retrieval.  Included are the
    target minimum (open purple circle at the average of three local
    minima) and the chosen minimum on the unshaded region (closed
    purple circle).  The GlobCurrent-drifter shared error fraction
    ($\lambda_N$) at the chosen minimum is also shown.}
  \label{fig04}
\end{figure}

\begin{multicols}{2}
Figure~\ref{fig04} depicts the absolute difference in LHS minus RHS of
the forecast and revcast autocovariance equations
(\ref{infers_autocov}).  Differences are shown for all candidate
values (i.e., true variance between zero and the variance of
drifters), but model solutions are of interest only where variance is
positive (unshaded region).  The target minimum (open purple circle)
is the average of three available local minima (i.e., no minima are
associated with the extended forecast $E$).  Although this target
minimum is not accessible (on the unshaded region), the chosen true
variance solution is just to the right of this locus of three minima
and about the same distance from them as they are from each other.
This choice implies that at least one model variance estimate is zero.
Here, shading on the left in Fig.~\ref{fig04} corresponds to negative
shared true variance of the meridional current component (this is a
derived quantity that varies with $\lambda_N$).

Whereas target solutions on the unshaded region can be seen as a
reminder that models like (\ref{eq00}), (\ref{eqerr}), and
(\ref{infers}) are parsimonious \citep{Box_1979}, the tendency of
autocovariance minima to be found on the left side of Fig.~\ref{fig04}
may be the most important aspect of accommodating error
cross-correlation.  This first demonstration indicates that true
variance shared by GlobCurrent and drifters is as small as possible
(given that retrieved variance should be positive).  Visually, true
and drifter error variance are the abscissa lengths to the left and
right of the closed purple dot, respectively.  True variance is thus
smaller than drifter error variance when all collocations are
considered.
\end{multicols}

\vspace{-0.4cm}
\begin{table}[htb]
  \caption{Model parameters of the drifter ($I$) and GlobCurrent
    nowcast ($N$) zonal (U) and meridional (V) current components that
    are retrieved using 5310226 non-outlier collocations from the even
    years between 1993 and 2015 (cf. Fig.~\ref{fig02}).  Parameters
    include total standard deviation ($\sigma$), true standard
    deviation ($\sigma_t$), nowcast additive calibration ($\alpha_N$),
    multiplicative calibration ($\beta_N$), shared error fraction
    ($\lambda_N$), individual ($[1 - \lambda_N]^{\nicefrac{1}{2}}
    \sigma_I$ and $\sigma_N$) and total ($\sigma_I$ and $[\lambda_N^2
      \sigma_I^2 + \sigma_N^2]^{\nicefrac{1}{2}}$) error standard
    deviation as in (\ref{infers_strong}), signal correlation
    \citep{McColl_etal_2014}, and signal to noise ratio (SNR;
    \citealt{Gruber_etal_2016a}).  Standard deviation and additive
    calibration are given in ms$^{-1}$ and SNR is in dB.}
  \label{tab01}
  \begin{center}
    \begin{tabular}{|c|c|c|c|c|c|c|c|c|c|}
      \cline{2-10}
      \multicolumn{1}{c|}{} & $\sigma$ &                                $\sigma_t$ &                             $\alpha_N$ &              $\beta_N$ &            $\lambda_N$ & $\sigma_{indiv}$ & $\sigma_{total}$ & Corr & SNR \\
      \hline
      U$_I$             & 0.195 & \multirow{4}{*}{\makecell{U:\\0.127\\V:\\0.003}} &                             \multicolumn{3}{c|}{\multirow{2}{*}{}}                       &             0.100 &                             0.148 & 0.652 &  -1.3 \\
      \cline{1-2} \cline{7-10}
      V$_I$             & 0.159 &                                                  &                             \multicolumn{3}{c|}{}                                        &             0.107 &                             0.159 & 0.021 & -33.6 \\
      \cline{1-2} \cline{4-10}
      U$_N$             & 0.168 &                                                  &             -0.001                     & \multirow{2}{*}{0.843} & \multirow{2}{*}{0.546} &             0.100 &                             0.129 & 0.640 &  -1.6 \\
      \cline{1-2} \cline{4-4} \cline{7-10}
      V$_N$             & 0.130 &                                                  &              0.001                     &                        &                        &             0.097 &                             0.130 & 0.022 & -33.3 \\
      \hline
    \end{tabular}
  \end{center}
\end{table}
\vspace{-0.4cm}

\begin{multicols}{2}
Table~\ref{tab01} provides model parameters for the drifter (in situ)
and GlobCurrent (nowcast) zonal ($U$) and meridional ($V$) current
components.  We find that truth and error are of similar magnitude and
that GlobCurrent and drifters sample not only a shared truth but also
shared error.  However, this truth exists only in the zonal component
(0.127~ms$^{-1}$).  Negligible meridional amplitude (0.003~ms$^{-1}$)
corresponds with a solution at the border of the shaded region in
Fig.~\ref{fig04}.  The additive calibration of GlobCurrent
($\alpha_N$) is also negligible and multiplicative calibration
($\beta_N$) is prescribed by variance matching (Fig.~\ref{fig03}).
Evidently, GlobCurrent samples are within their autocovariance
envelope as the minimum correlation for this sample is 0.91 and 0.83
for the zonal and meridional current components, respectively.

We obtain most of the individual error terms in (\ref{infers}) and
(\ref{infers_strong}) from the model retrievals of unshared
(measurement error) variance (i.e., $\sigma_N^2, \sigma_F^2,
\sigma_E^2, \sigma_R^2,$ and $\sigma_S^2$).  The exception is
individual error for drifters ($[1 - \lambda_N] \epsilon_I$), which
follows from our definition of shared equation error
\citep{Kipnis_etal_1999}.  Diagnostic equations for shared and
unshared drifter error variance can be written as $\lambda_N
\sigma_I^2$ and $(1 - \lambda_N) \sigma_I^2$, respectively (i.e.,
assuming an even split of the covariance between equation error
$\lambda_N \epsilon_I$ and individual error $[1 - \lambda_N]
\epsilon_I$).  Because over 50\% of drifter error is shared by
GlobCurrent ($\lambda_N$), the percentage of total variance in
(\ref{infers_strong}) that is shared equation error ranges from 23\%
(GlobCurrent zonal component) to 55\% (drifter meridional component).

Individual and total error variance for the zonal and meridional
components are both high (Table~\ref{tab01}).  Calibration by variance
matching dictates that drifter and GlobCurrent correlation with truth
\citep{McColl_etal_2014} and SNR \citep{Gruber_etal_2016a} are roughly
the same by zonal or meridional component \citep{Su_etal_2014}.
Meridional noise dominates signal (SNR is -33dB) and even zonal noise
is larger than signal (SNR $<$ 0).  Note that SNR is calculated using
total error (i.e., both correlated and uncorrelated; third column from
the right in Table~\ref{tab01}).  A preliminary regional assessment
(not shown; \citealt{GlobCurrent_2017}) is consistent with previous
studies \citep{Johnson_etal_2007, Blockley_etal_2012, Sudre_etal_2013}
in highlighting that weak meridional SNR is a characteristic of the
equatorial regions.
\end{multicols}

\begin{figure}[htb]
  \centering
  \includegraphics[width=0.85\textwidth]{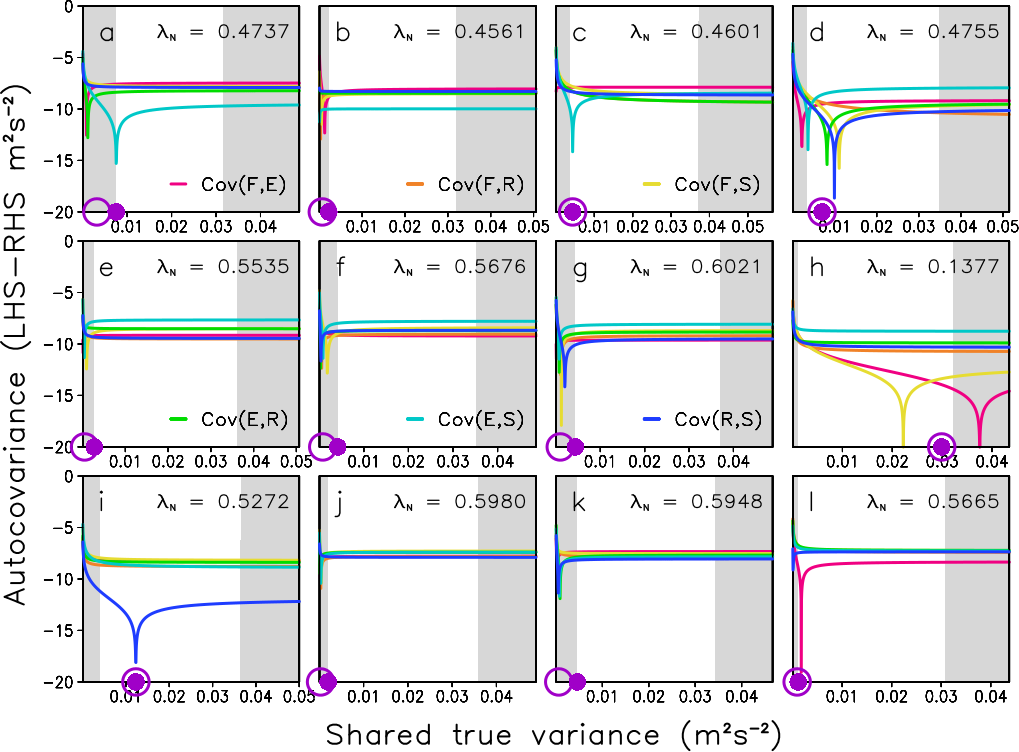}
  \caption{As in Fig.~\ref{fig04}, but only for collocations north of
    15$^\circ$N on day a)~30, b)~60, c)~90, d)~120, e)~150, f)~180,
    g)~210, h)~240, i)~270, j)~300, k)~330, and l)~360 of the year for
    even years between 1993 and 2015.}
  \label{fig05}
\end{figure}
\vspace{-0.2cm}

\begin{multicols}{2}
Figure~\ref{fig05} is a second demonstration that true variance shared
by GlobCurrent and drifters is small.  Parameters are retrieved as a
function of day of the year, and to isolate one high latitude seasonal
cycle, collocations north of 15$^\circ$N latitude are selected.  We
focus on 2385232 collocations of this northern region from even years
between 1993 and 2015 (i.e., 21\% of those available, using about 6000
collocations per day and applying variance matching and outlier
removal at daily intervals).  Figure~\ref{fig05} depicts solutions of
true variance for an arbitrary selection of 12~days, of which eight
are consistent with Fig.~\ref{fig04} insofar as the locus of
autocovariance minima (\ref{infers_autocov}) are at exceedingly small
true variance.  Only on day~240 (Fig.~\ref{fig05}h) is true variance
relatively large (as dictated by covariance involving F).  An
examination of all 364~days reveals a similar result: true variance is
as small as possible on 250~of 339~days (74\%).  No parameters are
estimated on 25~of 364~days (7\%) because no autocovariance minima are
found.

\begin{figure}[H]
  \centering
  \includegraphics[width=0.95\textwidth]{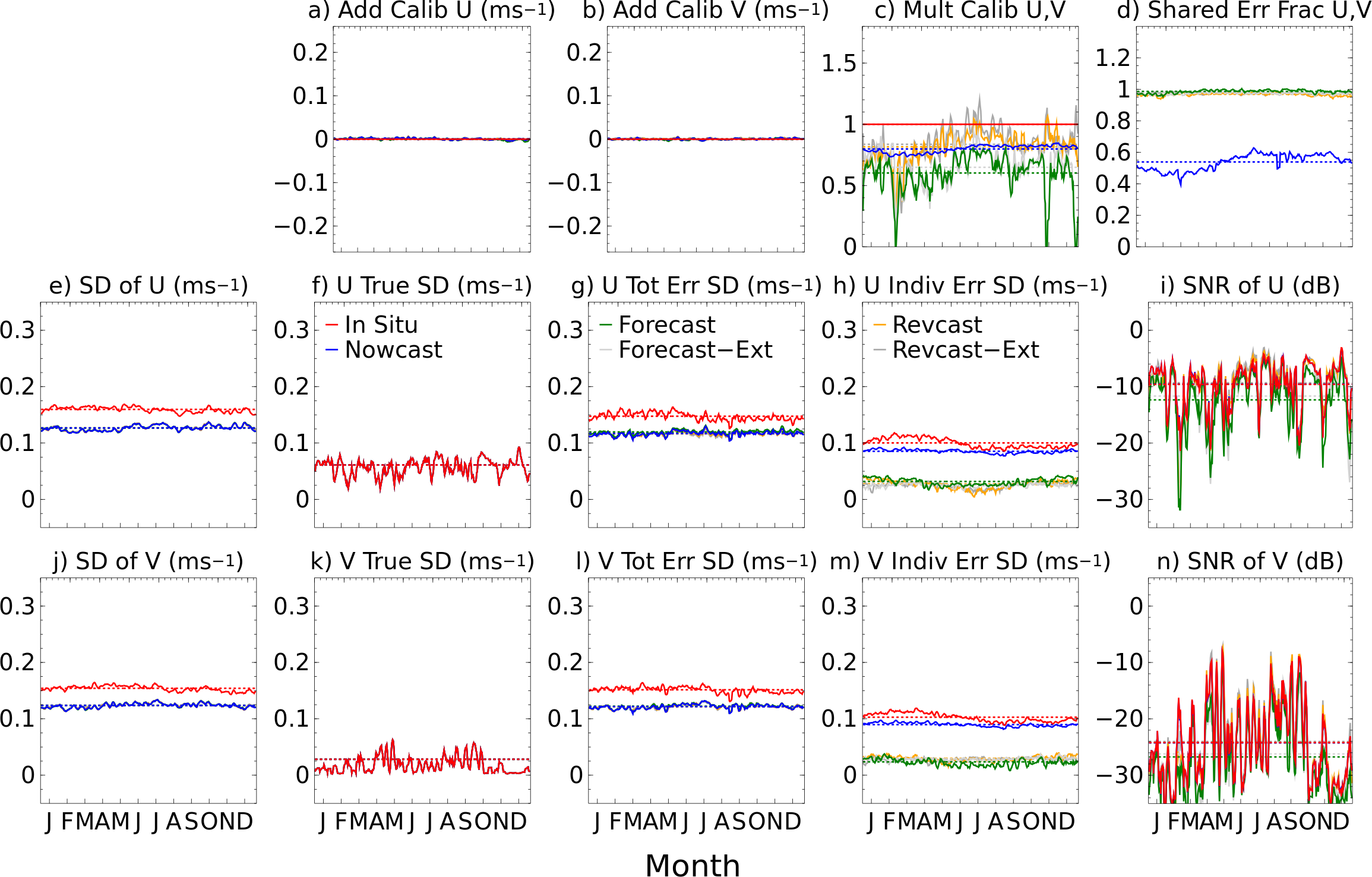}
  \caption{Retrieved model parameters as in Table~\ref{tab01}, but for
    339~days of the year using about 6000 collocations per day from
    north of 15$^\circ$N and from even years between 1993 and 2015.
    Shown are the drifter (in situ/red) and GlobCurrent (nowcast/blue,
    forecast/green, revcast/orange, and extended forecast/light grey
    and revcast/dark grey) retrievals of a)~zonal and b)~meridional
    additive calibration (ms$^{-1}$) and c)~multiplicative calibration
    and d)~shared error fraction for both zonal and meridional
    components, and e,j)~15-m current, f,k)~shared truth, g,l)~total
    error, and h,m)~individual error standard deviation (ms$^{-1}$),
    and i,n)~signal to noise ratio (dB) for the zonal and meridional
    components, respectively.  Solid lines are averages over five days
    and dashed lines are annual averages.}
  \label{fig06}
\end{figure}

Figure~\ref{fig06} depicts the Northern Hemisphere seasonal cycle by
five-day running means for the full set of INFERS model parameters.
There is an annual variation in the calibration and shared error
parameters (c,d) that can be explained by (e,j) GlobCurrent and
drifter variations being slightly more similar in amplitude toward the
end of the year than at the beginning (e.g., solid lines tend to
bracket the annual-average dashed lines in March and to be bracketed
by them in September).  Of course, this similarity is largely
superficial, based on a consistent retrieval throughout the year of
small shared truth in the zonal component (f; Fig.~\ref{fig05}), and
as in Table~\ref{tab01}, almost no signal in the meridional component
(k).

Drifter noise in Fig.~\ref{fig06} appears to be greater during spring
than fall whereas GlobCurrent signal (via seasonality in
multiplicative calibration) is the opposite.  As a result, signal to
noise ratio is higher for both GlobCurrent and drifters in late summer
compared to spring, even for the meridional current (despite its weak
signal).  A spatiotemporal refinement of this result (with specific
attention to the role of mixed layer depth) seems to be required.
This same seasonality in SNR is obtained for the forecast and revcast
samples, although via a different allocation of variance (i.e., with
total error being almost entirely defined by the GlobCurrent nowcast
error).  The range in multiplicative calibration (c) for the forecast
and revcast data is an a posteriori justification for retaining
separate calibrations in (\ref{infers}).  All NFERS GlobCurrent
samples again appear to be within their autocovariance envelope, as
the minimum correlation among all days of the year is 0.88 and 0.84
for the zonal and meridional current components, respectively.
\end{multicols}

\newpage

\begin{figure}[H]
  \centering
  \includegraphics[width=0.95\textwidth]{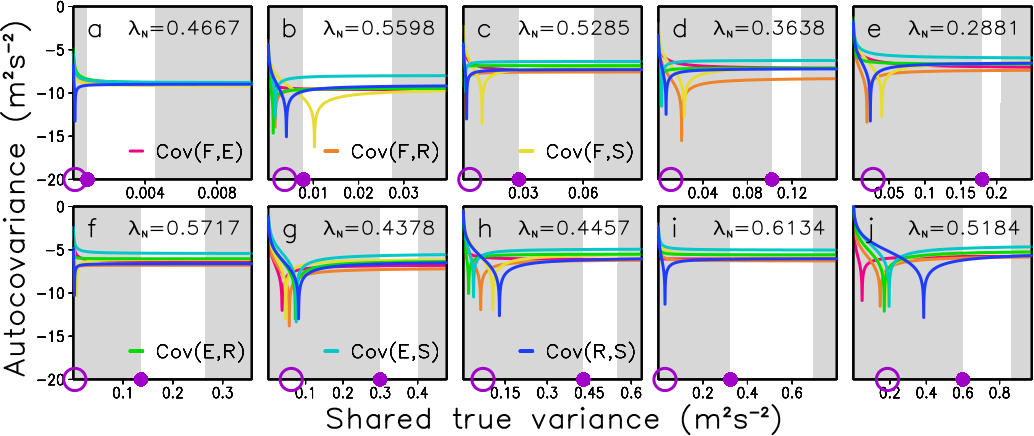}
  \caption{As in Fig.~\ref{fig04}, but for subsets of 500 collocations
    whose drifter speed is nearest to a)~0.1~ms${-1}$,
    b)~0.2~ms${-1}$, c)~0.3~ms${-1}$, d)~0.4~ms${-1}$,
    e)~0.5~ms${-1}$, f)~0.6~ms${-1}$, g)~0.7~ms${-1}$,
    h)~0.8~ms${-1}$, i)~0.9~ms${-1}$, and j)~1.0~ms$^{-1}$.
    Note that abscissa range varies with current speed.}
  \label{fig07}
\end{figure}
\vspace{-0.2cm}

\newpage
\begin{multicols}{2}
Figure~\ref{fig07} is the third demonstration that true variance
shared by GlobCurrent and drifters is small.  For a diagnosis of
model parameters as a function of drifter current speed, we again
apply variance matching and outlier removal \citep{Hubert_etal_2012}
as above, but to small groups of collocations.  \citet{Tolman_1998}
demonstrates that fine bin resolution (with sample sizes of O[100]) is
useful to avoid bias in covariance estimates.  Moreover,
\citet{Zwieback_etal_2012} recommend at least 500~collocations based
on idealized triple collocation simulations.  Solutions of true
variance are thus obtained over a finely resolved (0.01-ms$^{-1}$)
range in drifter speed using 500~collocations closest to each of
101~target speeds.  (This sampling requires less than 1\% of the
available collocations.)  Individual panels in Fig.~\ref{fig07} are
again consistent with Fig.~\ref{fig04} in that all 10~loci of
autocovariance minima (\ref{infers_autocov}) are at exceedingly small
true variance.  An examination of the 101~speed bins reveals that true
variance is as small as possible for 90~of 92~bins (98\%) and no
parameters are estimated for 9~of 101~bins (9\%) because no
autocovariance minima are found.
\end{multicols}

\begin{figure}[htb]
  \centering
  \includegraphics[width=\textwidth]{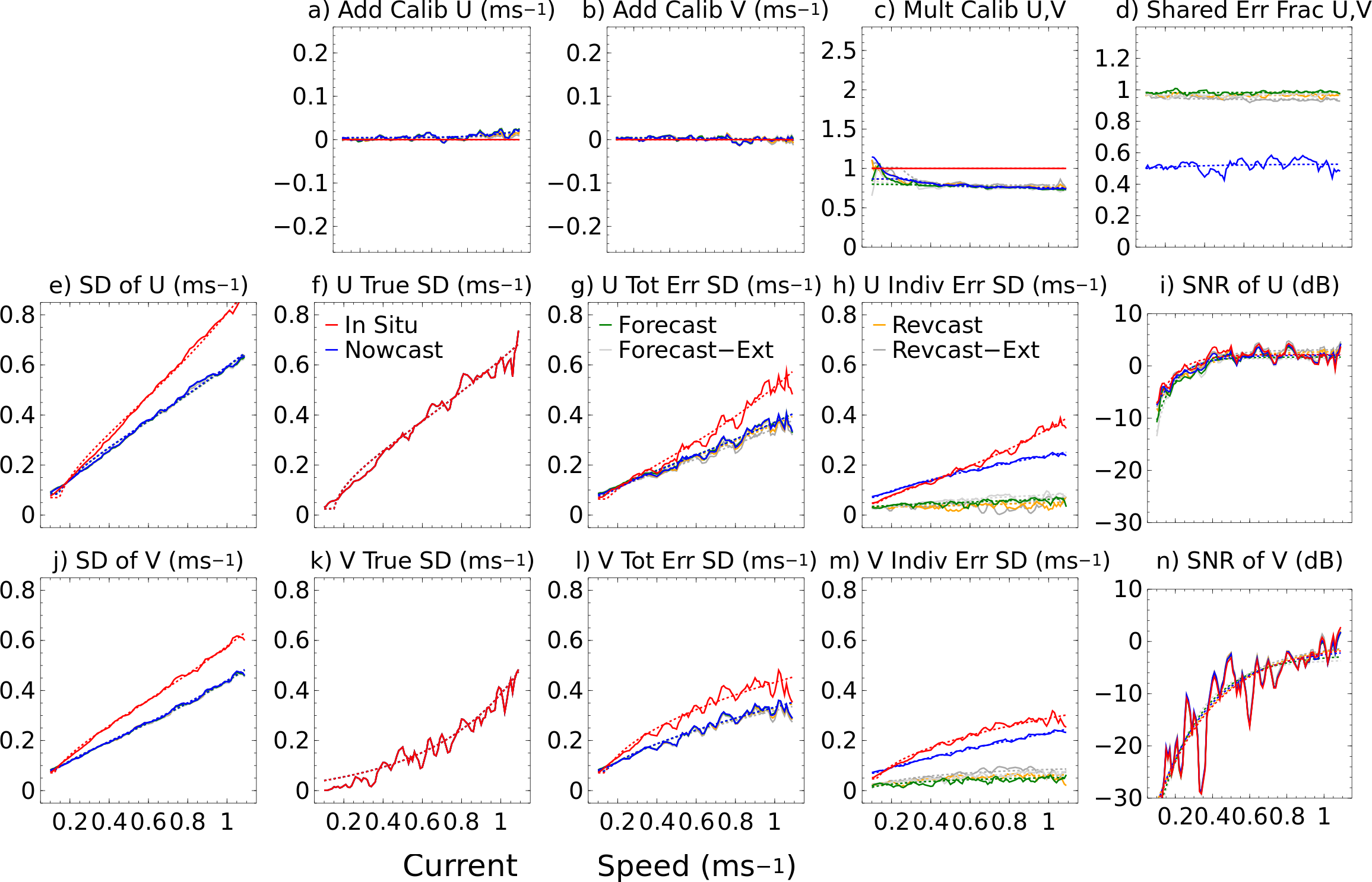}
  \caption{Model parameters as in Fig.~\ref{fig06}, but for
    92~subsets of 500 collocations whose drifter speed is nearest to
    target values between 0.1~ms$^{-1}$ and 1.1~ms$^{-1}$ at intervals
    of 0.01~ms$^{-1}$ (excluding 9~solutions for which no
    autocovariance minima were found).  Solid lines are averages of
    five adjacent intervals.  Dashed lines are best fits of the form
    $y(x) = a + b e^{cx}$ \citep{Jacquelin_2014}, but for c)
    multiplicative calibration, this fit ignores target values less
    than 0.3~ms$^{-1}$.}
  \label{fig08}
\end{figure}

\begin{multicols}{2}
Figure~\ref{fig08} illustrates the dependence of model parameters on
current speed.  There are weak trends in the calibration and shared
error parameters (a-d) and strong trends in most variance parameters
(e-n).  As in Table~\ref{tab01}, GlobCurrent-drifter shared error
fraction ($\lambda_N \approx 0.5$) is quite high, variance-matched
multiplicative calibration ($\beta_N$) is about 0.85 beyond
0.3~ms$^{-1}$, and additive calibration of GlobCurrent ($\alpha_N$) is
negligible.  Justification for our application of variance matching
thoughout this study (rather than assuming no GlobCurrent bias) is
that an upper bound on multiplicative bias, as given by reverse linear
regression, falls below one at large current speed (not shown).  In
turn, the need to address strong current underestimation (perhaps
locally in time and space, but at the resolution of the GlobCurrent
analysis) may continue to exist (cf.~\citealt{Rio_etal_2014}).

Errors in GlobCurrent samples separated by a day are basically the
same in Fig.~\ref{fig08}g,h,l,m.  The product of the forecast and
revcast shared error fraction parameters ($\lambda_F \lambda_E
\lambda_R \lambda_S$) is thus close to unity, which implies that
GlobCurrent error is being sampled within its autocovariance envelope.
In effect, this justifies the use of the extended forecast and revcast
samples in the INFERS model.  Among all 92~subsets, the minimum
correlation of combined truth and error (found at low speed between
$E$ and $S$) is 0.84 and 0.76 for the zonal and meridional current
components, respectively.

Figure~\ref{fig08}f,k reveals weak agreement between GlobCurrent and
drifters on a shared truth at low current speed, but more agreement at
higher current speed.  This is dictated in part by current speed
itself (Fig.~\ref{fig08}e,j), but the meridional component of drifter
error increases quickly with current speed (more so than the zonal
component) and the opposite is the case for true variance.  In
contrast to negative SNR for the zonal component in Table~\ref{tab01},
the GlobCurrent/drifter best fit SNR (Fig.~\ref{fig08}i dashed lines;
equivalent by variance matching) eventually exceed, but remain close
to, 0~dB from about 0.3~ms$^{-1}$.

This section constitutes an introduction to the INFERS model featuring
hundreds of parameter solutions.  Our experiments are thus enabled by
access to millions of drifter current estimates and a GlobCurrent
analysis that is about three orders of magnitude larger.  This is not
to say that 500 collocations is small.  In many contexts, including
ours, a few hundred collocations may be ample.  However, with the
freedom afforded by large datasets to identify a range of solutions
using appropriate instruments (cf.~\citealt{Kipnis_etal_2002}), comes
the opportunity to better characterize shared truth and error.  The
next section briefly explores shared truth as an updated measure of
agreement between variates and clarifies shared error as an updated
measure of dependence.

\section{Discussion}

It is sometimes the case in geophysics that only one truth (a
so-called genuine truth) is of interest.  Implicit in this concept is
the idea that truth carries no information about particular datasets,
which differ only in terms of their corresponding error, and this
error is intrinsic (i.e., defined without reference to another
dataset).  Implicit in the definition of shared truth, on the other
hand, is the idea that {\it if shared truth exists, then it contains
  information about an overlap in data supports} (see Appendix).
Beyond the scope of this paper, but notable within geophysics, are
formal inference theories that concern a conjunction of information
and the problem of aggregated opinion \citep{Tarantola_2005}.  Here,
it suffices to note that measurement models can provide a calibration
by linear mapping, and a validation by shared/unshared error, but they
can also provide a useful measure of agreement among datasets by
shared truth.


One documented application of shared truth is an assessment by
\citet{Bentamy_etal_2017} of various global ocean surface heat flux
analyses.  Using the INFERS model, Bentamy et al. experiment with
shared truth as a metric of competitive validation (see Appendix).
Following a recalibration of each gridded analysis to the same in situ
reference, they observe that in situ and analysis total error becomes
equal, whereas shared truth is invariant (their Table~2 thus provides
a standardized ranking).  This invariance of shared truth is a
property of many measurement models and may not be well known, perhaps
in part because shared truth itself is often undocumented.  To be
fair, all documented searches so far (including Bentamy et al.) assume
a fixed calibration rather than seeking true variance and calibration
together (cf.~Section~3).

\begin{figure}[H]
  \centering
  \includegraphics[width=0.8\textwidth]{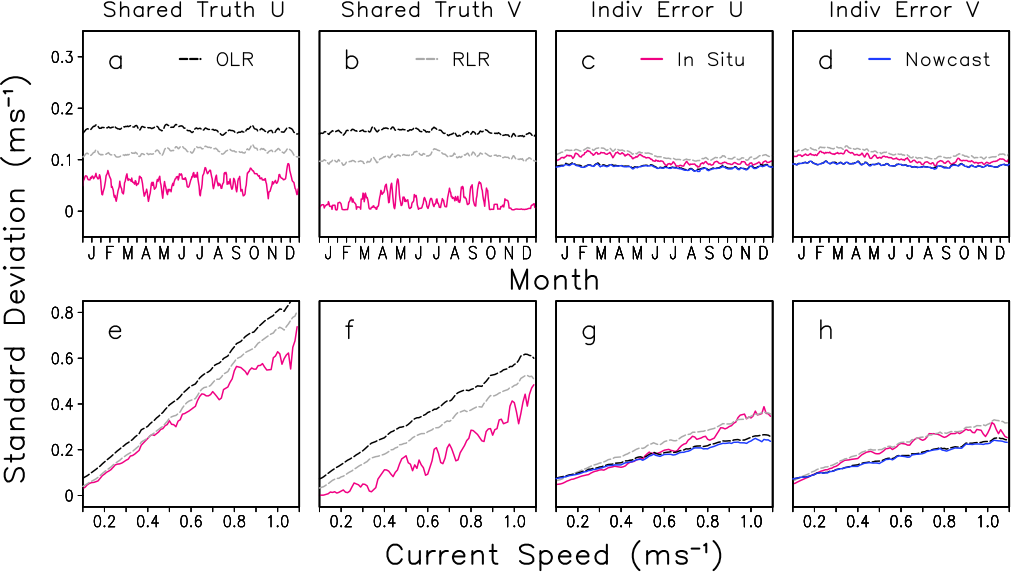}
  \caption{Shared truth (a,b,e,f) and individual error (c,d,g,h) as in
    Figs.~\ref{fig06} and~\ref{fig08} (f,k,h,m), but only for the
    drifter (in situ; red) and GlobCurrent (nowcast; blue)
    collocations.  Included are the corresponding ordinary (OLR;
    dashed black) and reverse (RLR; dashed grey) linear regression
    reference solutions.}
  \label{fig09}
\end{figure}

We propose that shared truth should have equal focus to error in
typical validation efforts.  Because INFERS introduces error
correlation into the errors-in-variables regression model, a good
comparison for INFERS is the full range of solutions consistent with
(\ref{eq00}), with familiar analytic solutions for ordinary (OLR) and
reverse (RLR) linear regression being appropriate references.
Solutions of the OLR and RLR models are identified by the method of
moments with either drifter error ($\epsilon_I$ for OLR) or
GlobCurrent error ($\epsilon_A$ for RLR) set to zero.  INFERS
estimates of truth and error from the previous section are placed
alongside these two reference solutions in Fig.~\ref{fig09}.  It is
notable that INFERS solutions of true standard deviation
(Fig.~\ref{fig09}a,b,e,f) are smallest.  This is remarkable because
the OLR and RLR references are understood to be the solutions that
bound the range of true variance (and multiplicative calibration or
regression slope) values that are consistent with the
errors-in-variables model (\ref{eq00}).

Further comparison between INFERS and the corresponding OLR and RLR
reference solutions permit an interpretation of the unshared
(measurement) errors that define much of the total error in this
study.  Figure~\ref{fig09}c,d,g,h reveals that the magnitude of OLR
error in GlobCurrent and RLR error in drifters appear to differ little
from the unshared error shown in Fig.~\ref{fig06}h,m and
Fig.~\ref{fig08}h,m.  As noted in Section~4, some ambiguity is
expected in a diagnostic estimate of drifter unshared error, but the
overlapping agreement in GlobCurrent unshared error (i.e., black
dashed and blue lines) is evident for all collocation divisions.
Whereas OLR and RLR impose separate assumptions on (\ref{eq00}) that
provide hypothetical bounds on uncorrelated error, in this study a
single model seems to provide both solutions.
\end{multicols}

\begin{multicols}{2}
Figure~\ref{fig09} reveals that total error in GlobCurrent and
drifters can be interpreted as a combination of respective RLR and OLR
upper limits in uncorrelated error.  Subject to the caveat that a
fixed calibration by variance matching allows more freedom for shared
error in our INFERS solutions, the reason that shared true variance
falls outside the OLR and RLR bounding reference solutions is because
not only can the INFERS model accommodate bounds on unshared
(measurement) error, as given by (\ref{eq00}), but shared (equation)
error is accommodated as well.


We conclude this initial characterization of model solutions by noting
that shared error offers an updated measure of error dependence.  It
is important to recognize that any decision to exclude shared error
from a measurement model, based on physical knowledge of the data
alone, can always be challenged.  In other words, {\it even if there
  is no apparent physical relationship between two datasets,
  independence of their errors should not be presumed without
  considering that the measurement model is only an approximation}
\citep{Box_1979}.  Thus, it may be appropriate to accommodate sharing
even if one cannot assume that shared error (or truth) exists.  More
specifically \citep{Fuller_1987}, if the model assumes that truth and
error are additive with a linearly related signal, as in (\ref{eq00}),
and this might not be strictly true of the data, then some form of
both equation and measurement error (\ref{eqerr}) or correlated and
uncorrelated error (\ref{infers}) should be included.

Equation error and correlated error are considered to be essentially
the same in this study, as we now demonstrate, but they are not
strictly the same error in general.  For instance,
\citet{Kipnis_etal_1999} allow for correlation in both equation error
and measurement error.  The Introduction acknowledges that GlobCurrent
and drifters also may share a component of measurement error.  This is
because many of the same drifters that are employed to refine the
CNES-CLS13 mean dynamic topography (MDT; \citealt{Rio_Hernandez_2004,
  Rio_etal_2014}) are employed above for validation.  Although INFERS
provides an estimate of error correlation that may include measurement
error, one option for demonstrating its interpretation as equation
error is a validation only after 2013.  Instead, we opt to replace the
CNES-CLS13 MDT in each GlobCurrent sample (NFERS) with a more
approximate GOCE-only MDT \citep{Rio_etal_2014}.  Drifter measurement
error is thus removed from GlobCurrent and the remaining error
correlation can be attributed entirely to equation error.
\end{multicols}

\begin{table}[htb]
  \caption{As in Table~\ref{tab01}, but for a
    measurement-error-independent comparison between GlobCurrent and
    drifters: GlobCurrent data exclude a velocity component associated
    with the CNES/CLS-2013 MDT and include instead a component
    associated with the GOCE-only geodetic MDT \citep{Rio_etal_2014}.
    Parameters of the drifter ($I$) and GlobCurrent nowcast ($N$)
    zonal (U) and meridional (V) current components are retrieved
    using 5280828 non-outlier collocations from the even years between
    1993 and 2015.}
  \label{tab02}
  \vspace{0.3cm}
  \begin{center}
    \begin{tabular}{|c|c|c|c|c|c|c|c|c|c|}
      \cline{2-10}
      \multicolumn{1}{c|}{} & $\sigma$ &                                $\sigma_t$ &                             $\alpha_N$ &              $\beta_N$ &            $\lambda_N$ & $\sigma_{indiv}$ & $\sigma_{total}$ & Corr & SNR \\
      \hline
      U$_I$             & 0.194 & \multirow{4}{*}{\makecell{U:\\0.119\\V:\\0.001}} &                             \multicolumn{3}{c|}{\multirow{2}{*}{}}                       &             0.106 &                             0.153 & 0.612 &  -2.2 \\
      \cline{1-2} \cline{7-10}
      V$_I$             & 0.158 &                                                  &                             \multicolumn{3}{c|}{}                                        &             0.110 &                             0.158 & 0.007 & -43.7 \\
      \cline{1-2} \cline{4-10}
      U$_N$             & 0.161 &                                                  &             -0.001                     & \multirow{2}{*}{0.818} & \multirow{2}{*}{0.517} &             0.101 &                             0.128 & 0.604 &  -2.4 \\
      \cline{1-2} \cline{4-4} \cline{7-10}
      V$_N$             & 0.127 &                                                  &              0.002                     &                        &                        &             0.097 &                             0.127 & 0.007 & -43.5 \\
      \hline
    \end{tabular}
  \end{center}
\end{table}

\begin{multicols}{2}
Table~\ref{tab02} provides a comparison between GlobCurrent (GOCE-only
MDT) and drifters based on 5280828 non-outlier collocations from the
even years between 1993 and 2015.  With some of the strongest current
components (i.e., most different in terms of MDT) again excluded as
outliers, true standard deviation in the zonal component decreases
slightly (0.127~ms$^{-1}$ to 0.119~ms$^{-1}$) for an MDT that lacks
drifter information.  Otherwise, the results of Table~\ref{tab01} are
reproduced, including small true variance in the meridional component
and a large shared error fraction ($\lambda_N = 0.517$).  Although
this is a measurement-error-independent comparison, it is nevertheless
clear that the two datasets are not independent.  Shared error
fraction in Table~\ref{tab01} is quite similar ($\lambda_N = 0.546$),
as is the percentage of total variance in (\ref{infers_strong}) that
is shared error, again ranging from 24\% for the GlobCurrent zonal
component to 52\% for the drifter meridional component.  The
implication is that there is little error correlation owing to drifter
measurement error in the CNES-CLS13 MDT.  There is instead large error
correlation owing to equation error.

\section{Conclusions}

This study provides an approach to the challenge of introducing and,
like any other model term, identifying cross-correlated error in
linear regression models such as (\ref{eq00}).  Subject to the caveat
that calibration is prescribed by variance matching (rather than being
jointly retrieved with shared true variance),
over 90\% of all attempts to retrieve model parameters for GlobCurrent
and drifters are successful.  Perhaps the more surprising aspect is
that, given two datasets, we require just a few additional samples of
the GlobCurrent analysis around the time of each drifter observation.
Compared to the frequency of these additional samples, necessary
confirmation of slow changes in the evolution of GlobCurrent and its
errors is also obtained.

Formulation of a new measurement model called INFERS (an acronym taken
from data sample names) is inspired by instrumental variable
regression \citep{Su_etal_2014} and specifically the triple
collocation approach \citep{Stoffelen_1998, Caires_Sterl_2003,
  Janssen_etal_2007, OCarroll_etal_2008, Vogelzang_etal_2011,
  Zwieback_etal_2012, McColl_etal_2014, Yilmaz_Crow_2014,
  Gruber_etal_2016a}.  Error propagation through the data samples is
modelled using a first-order autoregressive (AR-1) formula, except
that propagation begins with the collocated sample equations (IN),
which provide the cross-correlated error terms, and then includes a
temporally symmetric application of AR-1 to error autocorrelation in
the remaining equations (FERS).  The most direct model comparison is
to solutions of the linear errors-in-variables regression model
(\ref{eq00}) because this is the same model given by the collocated
sample equations (IN) if cross-correlated errors are ignored.  A
search for true variance in a limited parameter space of the INFERS
model (i.e., assuming the variance matching calibration) yields values
smaller than for any solution of (\ref{eq00}), as given by ordinary
(OLR) and reverse (RLR) linear regression bounds.  Over three quarters
of these model solutions (Fig.~\ref{fig09}) support the proposition
that truth and signal, as defined in the INFERS model, are small (see
also Table~2 of \citealt{Bentamy_etal_2017}).

If truth is considered a shared model variable just like error
(ignoring its unshared component), then shared true variance can be
considered a measure of agreement between GlobCurrent and drifters.
Inferences about measurement model approximations as well as overlaps
in data support are then possible.  While it would be unfortunate to
start with a true variance that is smaller than it actually is
(variance matching may yield such a bias), to start with a truth that
is larger than it actually is would likely be more worrisome.  This
study indicates that there is a potential to overstate the agreement
between GlobCurrent and drifters based on an inflated true variance in
the linear errors-in-variables model.  Like the triple collocation
model, OLR and RLR are just identified and necessarily lack a term for
cross-correlated error.  Because their solutions involve variance
budgets with fixed total variance, as in the LHS of
(\ref{infers_strong}), if total error is increased by introducing a
new error term (equation or correlated error), then true variance
decreases by the same amount.  Tables~1 and~2 reveal that roughly a
quarter to a half of the total variance in GlobCurrent and drifters is
shared error variance.  Presumably, shared error is a first order term
that could not be much larger and remain hidden.  Subsequent studies
are needed to confirm whether this masquerading of equation error as
truth is common for other datasets and whether it should be attributed
to limitations in the errors-in-variables model.  Solutions in the
full parameter space of the INFERS model (excluding prescribed
calibration) should also be sought.

Implications of measurement model assumptions (e.g., that truth and
error are additive with a linearly related signal) are discussed in
geophysics (e.g., \citealt{Janssen_etal_2007, Zwieback_etal_2016}),
and moreso in the statistical literature, where notions are
established regarding how to accommodate nonlinear signals in linear
regression by including equation error \citep{Fuller_1987,
  Carroll_Ruppert_1996}.  Furthermore, accommodation of equation error
and measurement error {\it correlation} is given in sophisticated
measurement models in epidemiology \citep{Kipnis_etal_1999,
  Kipnis_etal_2002}.  In turn, it appears that the opportunity to
simultaneously identify all parameters of such models can be taken up
in part by studies like this one that incorporate an experimental
sampling of large datasets.

A sufficient number of GlobCurrent samples is taken before and after
each collocation (as persistence forecasts and revcasts, respectively)
so that there are more covariance equations than model parameters.
Retrieval of the 17~INFERS model parameters employs variance matching
to first prescribe the calibration from GlobCurrent to drifters.  Six
autocovariance equations, involving the FERS samples, weakly constrain
shared true variance and the remaining 15~covariance equations are a
strong constraint on the remaining 15~unknown parameters.  Insofar as
true variance is weakly constrained, this study avoids a common
assumption that real data be cast in the form of a simple measurement
model.

Model solutions have been examined for collocation groups numbering
about six million (from eleven years), 6000 (on each day of the year
in the NH), and 500 (nearest drifter speeds at 0.01-ms$^{-1}$
intervals).  One must be cautious about groups of collocations both
large (if in situ error is autocorrelated) and small (if parameter
retrievals depend on individual collocations;
cf.~\citealt{Zwieback_etal_2012}).  However, for all these subsets,
SNR is near zero at best because the error in GlobCurrent and drifters
is high, while variance of the true current is low.  There are
indications that the preferentially low SNR of the meridional
component is a characteristic of equatorial regions
(cf.~\citealt{Johnson_etal_2007, Blockley_etal_2012,
  Sudre_etal_2013}).  The interpretation of large individual error is
also interesting in that the OLR and RLR reference bounds on
uncorrelated error are reached by both GlobCurrent and drifters.

The last experiment of the Discussion is perhaps the most relevant for
an interpretation of shared and unshared error in terms of equation
and measurement error, respectively.  A measurement-error-independent
comparison between GlobCurrent (using a GOCE-only MDT) and drifters
permits a diagnosis of just how large the correlation in equation
error may be.  There is little change in shared error fraction between
the two MDT experiments, which suggests that correlated error in other
comparisons of this study may be viewed as predominantly that of
equation error rather than measurement error (in spite of a drifter
error contribution to the CNES/CLS13 MDT).  Good correspondence
between equation error and correlated error provides further impetus
for a review of common model assumptions.

The so-called genuine truth is not viewed in this study as the same
true variable $t$ that appears in most measurement models.  The search
for a genuine ocean surface current is ongoing, however, and iterative
or comparative applications of a measurement model have a role to play
(e.g., \citealt{Bentamy_etal_2017}).  By analogy with efforts to
validate SST, surface current depth should be useful to distinguish
between a slower, quasi-balanced flow and interactions with the
atmosphere.  For example, both drifters and GlobCurrent may be good
references for balanced flow experiments at the equator
(cf.~\citealt{Chan_Shepherd_2014}) and at higher latitudes
(cf.~\citealt{Penven_etal_2014}).  High resolution analyses are
expected to grow in number, and while validation is not a prescription
for finding the genuine current, there is an opportunity to quantify
improvements in two or more datasets (or versions of a single dataset)
against one chosen reference dataset.  This study documents variations
in INFERS model parameters as a function of day of the year and
current speed, but a high latitude flow experiment may benefit from
distinctions between cyclonic and anticyclonic eddies, whereas an
equatorial experiment may opt to treat the zonal and meridional
components separately.  With a view to mapping model parameters in the
dimensions of large datasets, an important challenge involves
selecting subsets of collocations according to an informed physical
understanding.

This study is a contribution to efforts of the geophysical community
to construct high resolution ocean surface current analyses using
assimilative numerical models and a synergy of observations (this
issue).  Because observational coverage is sparse, especially over the
ocean and in early years, a topical question remains whether to
withhold reference observations from an analysis so as to later
perform an independent validation.  To respond to this question in the
negative would imply that the same observations should be allowed to
benefit both the construction of an analysis and its validation.  In
turn, shared signal and noise in observations and analyses need to be
considered and measurement models that accommodate both equation error
and measurement error are called for (cf.~\citealt{Caires_Sterl_2003,
  Gruber_etal_2016a}).  It appears that not only can a basis for
understanding shared signal and noise be found in literature, but a
year-on-year accumulation of geophysical observations and high
resolution data is permitting more freedom, and slightly less
parsimony, in experimental measurement modelling.

\section{Acknowledgements}

We are pleased to acknowledge an international effort over many years
to collect, assemble, and analyze altimetric and drifter observations,
as well as the support and discussions (commencing roughly in reverse
time) with Bash Toulany, Will Perrie, Graham Dunn, Tim Williams, Igor
Esau, Laurent Bertino, Ad Stoffelen, Abderrahim Bentamy, Svetla
Hristova-Veleva, Bryan Stiles, Zorana Jelenak, Mike Brennan, Luc
Fillion, Bridget Thomas, Hal Ritchie, and Mike Dowd.  The opportunity
to reflect on the comments of three reviewers has been invaluable in
promoting a clearer presentation.  The Julia language
\citep{Bezanson_etal_2017} and R package DetMCD \citep{Hubert_etal_2012}
are also acknowledged.  Funding for this work (again in
reverse time) was from the European Space Agency via the Nansen Center
and Ifremer (Data User Element's GlobCurrent and Support to Science
Element's Ocean Heat Flux projects, respectively).  The first author
was also supported by the U.S.\ National Aeronautics and Space
Administration via the National Oceanic and Atmospheric Administration
and University Corporation for Atmospheric Research (Hurricane Science
Research, Ocean Vector Winds, and Visiting Scientist programs,
respectively), and the Canadian Space Agency via Environment and
Climate Change Canada (Government Related Initiatives program).
\end{multicols}


\bibliographystyle{amets}
\bibliography{refer}

\begin{multicols}{2}
\section{Appendix}

Measurement models (defined below) are actively evolving in various
fields, with geophysical applications that may be unfamiliar or are
just beginning to have an impact.  The solution of such models is
called an inverse problem \citep{Tarantola_2005}, by contrast with
evolution equations for mass, motion, and constituents as a forward
model.  It should be noted that longstanding experience in the
geo-physical/biological/chemical communities with forward modelling
and with taking high resolution (so-called longitudinal) observations
provide the basis for estimating error autocorrelation (e.g., using
FERS).  A brief clarification of other concepts relevant to this study
is offered here as a complement to more formal definitions.  Online
sources (e.g., Wikipedia) also provide recent and useful collaborative
summaries.  Concepts relevant to this study include:

\begin{itemize}
\item Affine calibration: synonymous with a linear calibration by
  intercept ($\alpha_N$) and slope ($\beta_N$) parameters.  Adjustment
  of the nowcast data ($N$) by these parameters is a good test of the
  retrieval method, as the adjusted nowcast should be unbiased.
  Regardless of the method, however, it is important to note that no
  bias correction can fully address a mismatch in support.

\item Autoregressive (AR) parameterization: an established expression
  of information propagation; used here to encompass not just error
  autocorrelation in time or space but also error cross-correlation
  between two ocean current variates.  The first order (AR-1) form
  explored here is the simplest.

\item Competitive validation: evaluation of two or more datasets (or
  versions of a single dataset) against one chosen reference dataset,
  where the metric of success is shared true variance.  Even if linear
  calibration is postulated (as in this study, rather than estimated
  from a measurement model), removal of linear bias from one dataset
  has no impact on shared truth, but this is not so for error.  This
  approach was first attempted by \citet{Bentamy_etal_2017} in a
  comparison of heat flux estimates.

\item Footprint: target area (e.g., at the ocean surface) that
  contributes to radiation received by a satellite sensor during an
  imaging interval.  Unless it is possible to combine views of the
  same target area to synthesize higher resolution, the footprint
  often defines a support scale lower bound.

\item Instrumental variable: additional data is often required when
  the measurement model has too many unknown parameters to estimate.
  A conventional instrument, following \citet{Fuller_2006}, is a
  variable that is taken to be correlated with truth but not with
  error.  The forecast and revcast (FERS) lagged variables, by
  comparison, involve correlation of both truth and error, but this is
  accommodated by their model equations.  As instruments, FERS play
  the required role of facilitating the identification of all model
  parameters.

\item Measurement model: measurement {\it error} models accommodate
  error in all sources of information [i.e., both in the calibrated
    and uncalibrated data; this accommodation is known as
    \citep{Fuller_2006} an approach to errors in variables in
    econometrics and observation error or measurement error in other
    fields].  There is no intended distinction between a measurement
  model and measurement error model.  The sole rationale for omitting
  the term ``error'' is that a more balanced focus on truth and error
  can be anticipated.  In other words, a regression model is
  effectively a truth model as much as it is an error model.  However,
  only if it is possible to claim that a model does not lack any broad
  category of error (i.e., equation error or correlated error), does
  it seem justifiable to explore inferences based on truth.

\item Parsimony: synonymous with simplicity, especially in reference
  to measurement models that minimize the number of parameters to be
  identified.  That is, non-technical definitions apply (e.g., to a
  careful collection or use of data with minimal extra assumptions).

\item Shared variance: synonymous with correlation and involving a
  term that appears in more than one of the measurement model
  equations of interest (possibly multiplied by a parameter).  The
  concept of sharing applies to both truth and error.  It is central
  to the idea that there can be multiple truths, with each containing
  information about overlapping data supports, and that measurement
  model assumptions should be considered when determining statistical
  independence.  It should be noted that standard metrics, including
  the coefficient of determination or percentage of explained
  variance, correlation with truth \citep{McColl_etal_2014}, and SNR
  \citep{Gruber_etal_2016a} are all subject to interpretation in terms
  of shared variance.

\item Strong constraint: as an example, many equations of the
  GlobCurrent and drifter covariance matrix (\ref{infers_strong}) are
  satisfied exactly as part of any measurement model solution
  (cf.~weak constraint).

\item Support: a characterization of the type (e.g., range or quality)
  of information that a given platform or instrument is sensitive to.
  Often this is with reference to spatial and temporal scales that can
  be resolved, but any information sensitivity can be included, which
  implies that such information may exist as truth or perhaps as
  equation error, according to the measurement model.

\item Synergy: an approach to combining information such that the
  whole is more valuable and informative than the sum of individual
  contributions.  Measurement modelling is an unlikely tool to
  prescribe how synergy could be achieved, but may permit the
  quantitative exploration of both individual contributions and
  informed attempts to combine information.

\item Triple collocation: following \citet{McColl_etal_2014}, the
  model parameters sought are uncorrelated error variance of three
  independent datasets, and with one dataset as a reference, additive
  and multiplicative calibration of the other two.  Following
  \citet{Stoffelen_1998}, this measurement model implicitly includes
  cross-correlated error (e.g., representativeness error) because
  three different sources of information invariably have three
  different supports, so at least between two information sources with
  broader support (e.g., higher resolution), error cross-correlation
  would be expected.

\item True variance estimation: curves of the LHS-RHS of the
  autocovariance equations (\ref{infers_autocov}) are each
  characterized by a single localized minimum and flatness elsewhere
  in the range of zero to $Var(I)$.  The present study treats each
  available minimum as an equally good estimate of shared true
  variance and their average is taken.  This is in contrast to a
  global minimum sought using the average of all such curves.
  However, minima are often not overlapping so the global minimum is
  effectively a selection among one of the six possible minima.  This
  implies a reliance on the accuracy of each curve in representing its
  own (very small) minimum value, which might be ill advised.

\item Weak constraint: as an example, the autocovariance equations
  provide different target estimates of shared true variance that
  cannot all be satisfied simultaneously; a solution close to the
  center of the ensemble is thus adopted (cf.~strong constraint).

\end{itemize}
\end{multicols}

\begin{center}
  \includegraphics[width=\textwidth]{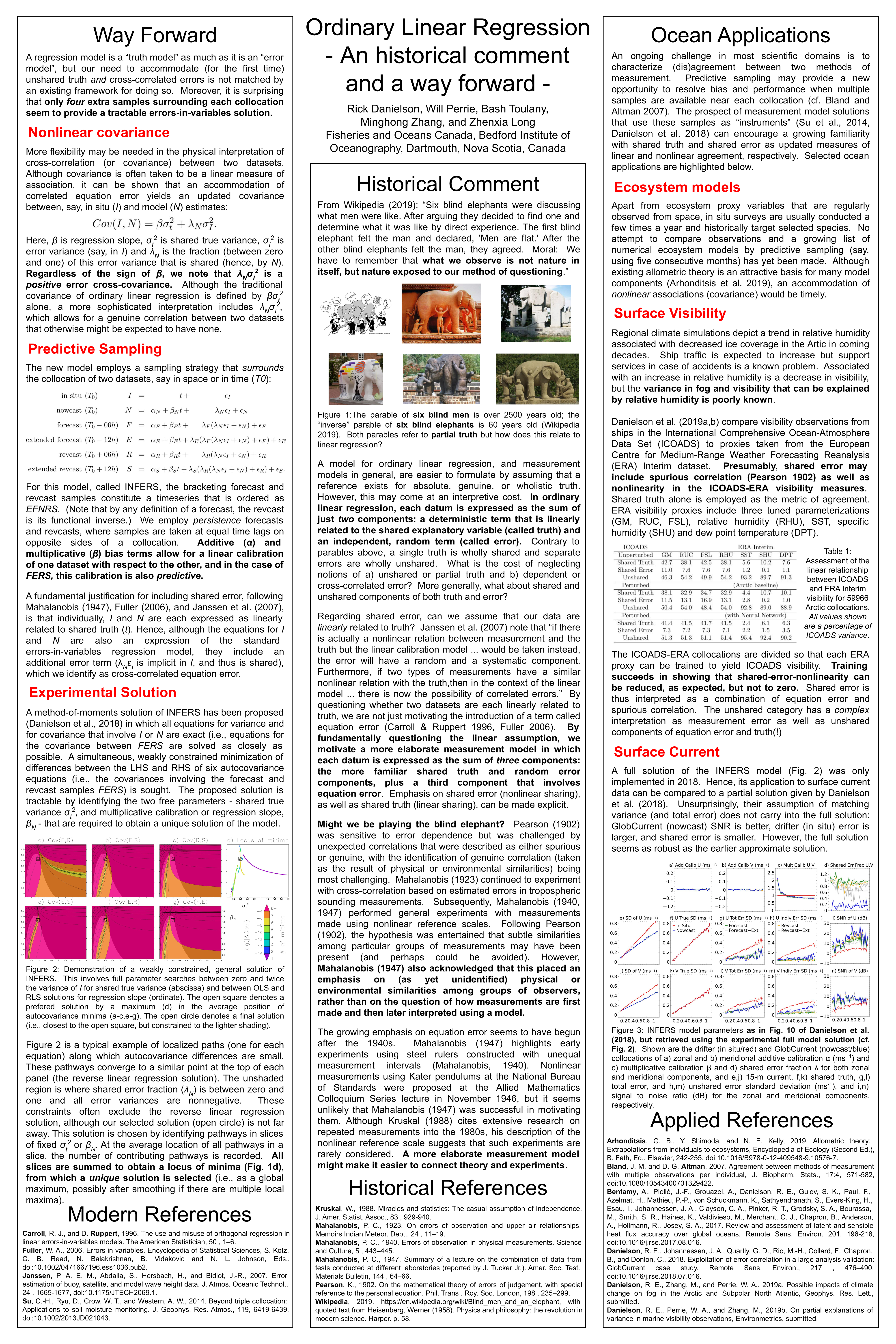}
\end{center}
\end{document}